\documentclass[12pt]{article}
\usepackage{latexsym}
\usepackage{a4,amssymb,amsmath}
\usepackage[dvips]{graphicx}
\begin{document}

\def\sh {\mathop {\rm sh} \nolimits}
\def\ch {\mathop {\rm ch} \nolimits}
\def\var {\mathop {\rm var}}

\centerline {\bf {CANONICAL QUANTIZATION of TWO-DIMENSIONAL GRAVITY}}

\centerline {}
\centerline {\it {S.N.Vergeles }}
\centerline {}
{ \small {\it {
\centerline {Landau Institute of Theoretical Physics, Russian Academy
of Sciences}

\centerline {Chernogolovka, Moscow Region, 142432 Russia.
e-mail: vergeles@itp.ac.ru}}}}
\centerline{}

\vspace {3mm}
\centerline{}

  \parbox [b] {145mm} {
 {\hspace {5mm} { \small { A canonical quantization of
two-dimensional gravity minimally coupled to real scalar and
spinor Majorana fields is presented. The physical state
space of the theory is completely described and calculations
are also made of the average value of the metric tensor
relative to states close to the ground state.}}}}


\centerline {}
\centerline {\bf {1. Introduction}}
\centerline {}

The quantum theory of gravity in four-dimensional Minkowski
space-time encounters fundamental difficulties which have
not yet been surmounted. These difficulties can
be conditionally divided into conceptual and computational.
The main conceptual problem is that the Hamiltonian is a
linear combination of first-class constraints. This fact
makes the role of time in gravity unclear. The main
computational problem is the nonrenormalizability of the
theory. The indicated difficulties are closely bound. For
example, depending on the computational procedure, the
constraint algebra may or may not contain an anomalous
contribution (central charge). The presence or absence of an
anomaly in the first-class constraint algebra has a decisive
influence on the quantization procedure and the ensuing
physical picture.

These fundamental problems can be successfully resolved
using relatively simple models of generally covariant
theories in two-dimensional space-time. These models
particularly include two-dimensional gravity models, both
pure and interacting with matter, and also two-dimensional
string models (see, for example [1-6] and the literature
cited therein).

In the present paper we spend a canonical quantization of
two-dimensional gravity minimally coupled to real scalar and
spinor Majorana fields. All the constructions and
calculations are given before the final result is obtained.
The physical states of the theory are completely described.
The complete state space has similar properties to the
multidimensional Fock
 space in which boson and fermion operators are acting. The
 average values of the metric tensor relative to states
 close to the ground state are calculated.

The progress reached at the construction of a
two-dimensional quantum theory of gravity is based on two
ideas. These ideas will be formulated below after the
introduction of the necessary notations.

Let's guess that the space-time is topologically equivalent
to a two-dimensional cylinder. The time coordinate \ $t$ \
varies between minus infinity and plus infinity while the
spatial coordinate \ $ \sigma $ \ varies between \ $ 0 $ \
and \ $2\pi $. All the functions are periodic with respect
to the coordinate \ $ \sigma $. The set of coordinates \ $
(t, \, \sigma) $ \ is designated also as \ $ \{x ^ {\mu} \,
\} $. The metric tensor in space-time is
 denoted by \ $g_{\mu\nu}$, so that the square of the interval is written as

$$
 ds^2=g_{00}\,dt^2+g_{11}\,d\sigma^2+2g_{01}\,dt\,d\sigma\,.
 \eqno (1.1)
$$
Most of the formulas and notations in the Introduction are
taken from [2]. The metric tensor is then parameterized as
follows:
$$
g _ {\mu\nu} =e ^ {2\rho} \, \left (
\begin {array} {cc}
u^2-v^2 & v \\
 v & -1
\end {array}
\right) \,, \ \ \ g\equiv det \, g _ {\mu\nu} = -u^2 \, e ^ {4\rho}\,. \eqno (1.2)
$$

Let \ $i, \, j=0, \, 1 $ \ and \ $ \eta _ {ij} =diag \, (1,
-1) $. Let's introduce the orthonormalized basis \ $ \{e ^
{\mu} _i \,\} $ so that
$$
g _ {\mu\nu} \, e ^ {\mu} _ie ^ {\nu} _j =\eta _ {ij}\,.
\eqno (1.3)
$$
To be specific we take
$$
e ^ {\mu} _0 =\frac {1} {u} \, e ^ {-\rho} \, \left (
\begin {array} {c}
1 \\
v
\end {array} \right) \,, \ \ \
e ^ {\mu} _1 =\left (
\begin {array} {c}
0 \\
e ^ {-\rho}
\end {array} \right)\,. \eqno (1.4)
$$
The dyad (zweibein) \ $ \{e^i _ {\mu} \,\} $ \ is uniquely
determinated
 by the equations \ $e^i _{\mu} e ^ {\nu} _i = $ $ \delta ^
 {\nu} _ {\mu} \Longleftrightarrow $ $e^i _ {\mu} e ^ {\mu}
 _j =\delta^i_j $. Taking into account (1.4), we have
$$
e^0 _ {\mu} = \left (
\begin {array} {c}
u \, e ^ {\rho} \\
0
\end {array} \right) \,, \ \ \
e^1 _ {\mu} =e ^ {\rho} \, \left (
\begin {array} {c}
-v \\
1
\end {array} \right)\,. \eqno (1.5)
$$

Let's consider the action
$$
 S=\int\,dt\,\int_0^{2\pi}\,d\sigma\,\sqrt{-g}\,
 \left\{\frac{1}{4\pi\,G}\,(\eta\,R-2\lambda\,)+
 \frac{1}{2}\,g^{\mu\nu}\,\partial_{\mu}f\,\partial_{\nu}f+
 \frac{i}{2}\,e^{\mu}_j\,\bar{\psi}\gamma^j{\cal D} _ {\mu} \, \psi \,
\right \}\,. \eqno (1.6)
$$
Here \ $G $ \ is the gravitational constant, \ $ \lambda $ \
is the cosmological constant, \ $R $ \ is the scalar
space-time curvature, \ $ \eta $\ and \ $f $ \ are the real
scalar fields, \ $ \psi $ \ is the two-component spinor
Majorana field and \ $ \{\gamma^i \, \}$ \ are
two-dimensional Dirac matrices. Further we assume that
$$
\gamma^0 =\left(
\begin {array} {cc}
0 & 1 \\ 1 & 0
\end {array} \right) \,, \ \
\gamma^1 =\left (
\begin {array} {cc}
0 & -1 \\ 1 & 0
\end {array} \right) \,, \ \
\gamma^5\equiv\gamma^0\gamma^1 =\left (
\begin {array} {cc}
1 & 0 \\ 0 & -1
\end {array} \right)\,. \eqno (1.7)
$$

The Majorana nature of the spinor field means that \ $ \psi
=\gamma^0 \,\bar {\psi} ^t $. The superscript \ $t $ \ means
transposition. In our case we have:
$$
\psi =\left (
\begin {array} {c}
\phi \\
\chi
\end {array} \right) \,, \ \ \ \phi =\phi ^ {\dag} \,, \ \ \
\chi =\chi ^ {\dag}\,. \eqno (1.8)
$$
The covariant differentiation operation of a spinor field is
defined according to the formula
$$
{ \cal D}_{\mu}\psi=\left(\,\frac{\partial}{\partial x ^ {\mu}} +
 \frac{1}{2}\,\omega_{ij\mu}\,\sigma^{ij}\,\right)\,\psi\,,
 \ \ \sigma^{ij}=\frac{1}{4}\,[\gamma^i,\,\gamma^j\,]\,.
 \eqno (1.9)
$$
The connection form \ $ \omega _ {ij\mu} $ \ is obtained
uniquely from the equation
$$
d\omega^i +\omega^i_j\wedge\omega^j=0 \,
$$
where \ $ \omega^i\equiv e^i _ {\mu} \, dx ^ {\mu} $. Thus we find
$$
 \omega_{01}=\left[u'+u\,\rho'-\frac{v}{u}\,(\dot{\rho}+\rho'\,v+v'\,)
 \,\right]\,dt+
 \frac{1}{u}\,(\dot{\rho}+\rho'\,v+v'\,)\,d\sigma\,. \eqno
 (1.10)
$$
Hereinafter the dot and prime above denote the partial derivatives
\ $ (\partial/\partial t) $ \ and \ $ (\partial/\partial\sigma) $,
respectively. Using the Cartan structure equation, we easily establish
 the formula
$$
\sqrt {-g} \, R \, dt\wedge d\sigma=2 \, d\omega ^ {01}\,. \eqno (1.11)
$$

As the fields \ $ \phi $ \ and \ $ \chi $ \ in (1.8) are real and
belong to Grassmann algebra, we have
$$
\phi (x) \, \phi (x) =0 \,, \qquad \ \ \chi (x) \, \chi (x) =0\,. \eqno (1.12)
$$

Using (1.12), we can make the substitution \ $ {\cal D}_
{\mu}\psi\rightarrow(\partial/\partial x ^ {\mu}) \, \psi $
in (1.6). Therefore the fermion part of the action is
proportional to the expression
$$
e^{\rho}\,\{\phi\,\dot{\phi}+(u+v)\,\phi\,\phi'+\chi\,\dot{\chi}-
(u-v) \, \chi \,\chi ' \}\,.
$$
In this last expression the factor \ $e^{\rho}$ \ may be
absorbed by substituting
$$
\phi\rightarrow e ^ {-\rho/2} \, \phi \,, \ \
\chi\rightarrow e ^ {-\rho/2} \, \chi\,.
$$
On account of (1.12), no additional derivative of the field
\ $ \rho $ appear in the action when this substitution is
made. Thus, the Lagrangian of the considered system has the
form
$$
{\cal L}=\int\,d\sigma\,\left\{\,\frac{1}{2\pi\,G}\,
  \left[\,u^{-1}\dot{\eta}\,(\dot{\rho}+v\,\rho'+v'\,)+
 \eta'\,\left(\,\frac{v}{u}\,(\dot{\rho}+v\,\rho'+v'\,)-
 u'-u\,\rho'\,\right)-\lambda\,u\,e^{2\rho}\,\right]+\right.
$$
$$
\left.
 +\frac{1}{2u}\,[\dot{f}^2+2v\,\dot{f}\,f'-(u^2-v^2)\,f^{\prime
 2}\,]+\frac{i}{2}\,[\phi\,\dot{\phi}+(u+v)\,\phi\,\phi'+
\chi \,\dot{\chi} -
(u-v) \, \chi \,\chi ' \,] \,\right \}\,. \eqno (1.13)
$$

Let's denote by \ $ \pi _ {\eta} \, \ \pi _ {\rho} $ \ and \
$ \pi $ \ the fields canonically conjugate to the fields \ $
\eta \,, \ \rho $ \ and \ $f $, respectively. The fields \
$u $ \ and \ $v $ \ in (1.12) are Lagrange multipliers. We
obtain the Hamiltonian of the system (1.13) by a standard
procedure:
$$
{ \cal H} = \int \, d\sigma \, (u \, {\cal E} +v \, {\cal P}
\,) \,,
$$
$$
{ \cal E} =2\pi \, G \,\pi _ {\eta} \pi _ {\rho} +
 \frac{1}{2\pi\,G}\,[-(\eta''-\eta'\,\rho')+
\lambda \, e ^ {2\rho} \,] +
\frac {1} {2} \, ( \pi^2+f ^ {\prime 2}\,) +
 \frac{i}{2}\,(-\phi\,\phi'+\chi\,\chi'\,)\,,
$$
$$
{ \cal P}=-\left(\pi_{\eta}\,\eta'+\pi_{\rho}\,\rho'+\pi\,f'+
\frac {i} {2} \, \phi \,\phi ' +
\frac {i} {2} \, \chi \,\chi ' \,\right)\,. \eqno (1.14)
$$
Let's make the following canonical transformation of variables:
$$
 \lambda\,r^0=-\frac{1}{2}\,\sqrt{\frac{\lambda}{4\pi\,G}}\,
e ^ {-\rho} \, ( 2\eta ' \,\ch \sum-4\pi \, G \,\pi _ {\rho}
\, \sh\sum \,) \,,
$$
$$
 \lambda\,r^1=\frac{1}{2}\,\sqrt{\frac{\lambda}{4\pi\,G}}\,
e ^ {-\rho} \, ( 4\pi \, G \,\pi _ {\rho} \, \ch \sum-2\eta
'
\,\sh \sum \,) \,,
$$
$$
\pi_0- {r ^ {1}} ^ {\prime} =
 \sqrt{\frac{\lambda}{\pi\,G}}\,e^{\rho}\,\sh\,\sum\,, \ \
 \pi_1+{r^{0}}^{\prime}=-\sqrt{\frac{\lambda}{\pi\,G}}\,e^{\rho}\,\ch\,\sum\,.
     \eqno (1.15)
$$
Here
$$
 \sum(\sigma)=2\pi\,G\,\int_0^{\sigma}\,d\tilde{\sigma}\,
\pi _ {\eta} (\tilde {\sigma})\,.
$$
The variables describing the matter remain unchanged. In the
new variables the Hamiltonian (1.14) becomes
$$
{ \cal E}=\frac{1}{2}\,[-(\pi_0^2+({r^0}^{\prime})^2\,)+ (
\pi_1^2 + ({r^1} ^ {\prime}) ^2 \,) \,] +
\frac {1} {2} \, [ \pi^2+f^{\prime 2} +i \, (-\phi \,\phi' +
\chi \,\chi') \,] \,,
$$
$$
{\cal P}=-(\pi_0\,{r^0}^{\prime}+\pi_1\,{r^1}^{\prime}\,)- (
\pi\,f'+\frac{i}{2}\,\phi\,\phi'+\frac{i}{2}\,\chi\,\chi'\,)\,.
   \eqno(1.16)
$$

So far the analysis was classical. To start quantization of
the system we must first begin by defining the simultaneous
permutation relations for canonically conjugate variables.
In our case, we have:
$$
[
r^0(\sigma),\,\pi_0(\sigma')\,]=[r^1(\sigma),\,\pi_1(\sigma')\,]=
[ f(\sigma),\,\pi(\sigma')\,]=i\,\delta(\sigma-\sigma ')\,.
\eqno (1.17)
$$
For the fermion degrees of freedom we have the anticommutation relations
$$
 \{\phi(\sigma),\,\phi(\sigma')\,\}=\{\chi(\sigma),\,\chi(\sigma')\,\}=
\delta (\sigma-\sigma ')\,. \eqno (1.18)
$$
All the other commutators or anticommutators of the
fundamental fields \ $r^a \,, \ \pi_a \,, \ f \,, \ \pi \,,
\ \psi $ \ are equal to zero. It is easy to check the
Heisenberg equations \ $i \,\dot {\cal O} = [{\cal O}, \,
{\cal H} \,] $, obtained using commutation relations (1.17),
(1.18) are the same as the Lagrange equations. Here \ $
{\cal O} $ \ is any operator.

Since the fields \ $u $ \ and \ $v $ \ in (1.14) are
Lagrange multipliers, the quantities (1.16) are constraints.
Within the framework of classical consideration they are the
first-class constraints. However, it is well known that as a
result of quantization, anomaly or central charge may
appear in this system: the algebra of simultaneous
commutators of \ $ {\cal E} $ \ and \ $ {\cal P} $ \
contains a central charge. The existence of a central charge
in the constraint algebra radically complicates the
quantization problem. In particular, the system (1.16) -
(1.18) can appear inconsistent.

Recently an anomaly-free approach to the quantization of the
system (1.16)-(1.18) has been proposed in various studies
[1-4, 6]. In this approach no central charge is present in
the quantum algebra of the quantities (1.16). It means, that
all the operators (1.16) can be treated as first-class
constraints in the Dirac sense. This new approach is applied
in the present study.

The idea of a new approach to quantization has arisen when
studying a model of pure gravity. This model is obtained
from the model (1.16) by eliminating the second terms on the
right-hand sides of the system (1.16). In [1-4, 6] it was
shown that in pure gravity theory the central charge is zero
if the scalar product is positively defined in the entire
state space. The reason for this phenomenon is that in the
new approach the operator ordering procedure in the values \
$ {\cal E} $ \ and \ $ {\cal P} $ \ differs radically from
the ordering in traditional quantization.

Now we shall formulate the assumptions on the basis of which
the new quantization method is developed.

1) The entire state space \ $H_C $, in which the fundamental
operator fields \ $r^a \,, \ \pi_a \,, \ f \,, \ \pi \,, \
\psi $ act, is supplied with a positively defined scalar
product. No indefinite metric is present in the \ $H_C $ \
space.

To formulate the next assumption, we denote the set of
operators (1.16) by \ $L $ \ and the set of all the material
fields $f \,, \ \psi $ by $ \psi $. Let's eliminate from the
operators \ $L $ \ the degrees of freedom describing the
material fields and denote the set of operators thus
obtained by \ $L^{(0)} $. Hence the operators \ $L ^ {(0)}
$ \ only determine the dynamics of the gravitational degrees
of freedom and
$$
[ L^{(0)}, \, \psi \,] = 0\,. \eqno (1.19)
$$

2) In the theory (1.16) there exists a unitary
transformation \ $U $ \ such that
$$
U \, L^{(0)} \, U^{\dag} =L \,. \eqno (1.20)
$$

As a result of (1.19) and (1.20) the fields
$$
\Psi=U \,\psi \, U^{\dag} \eqno (1.21)
$$
commute with all the operators \ $L $.

Let's explain the important role of the last assumption in
the quantization of the considered system. Let's postulate
that in the theory (1.16) there is a state \ $ | \, 0
\,\rangle $ which annuls all operators \ $L ^ {(0)} $ \ and
all the annihilation operators of the fields \ $ \psi $.
According to the reasoning put forward above this is
possible. Then the state \ $U \, | \, 0 \,\rangle $ \ is
annulled by all the operators
 \ $L $ \ and all the annihilation operators of the fields \
 $ \Psi $. The physical space of the states annulling all
 the operators \ $L $, is created from the ground state \ $U
 \, | \, 0 \,\rangle $ \ using the creation operators of the
 fields \ $ \Psi $. Thus the problem of quantization of the
 system (1.16) - (1.18) is solved completely.

In the second assumption, the properties of the unitary
transformation \ $U $ \ of interest to us are only described
in broad the contour. Subsequently this unitary
transformation is constructed explicitly for the model of
two-dimensional gravity studied here. The equivalent of the
formula (1.20) then has a more complex form. Nevertheless,
fairly good progress can be made in the calculations by
using the constructed unitary transformation.

\centerline {}
\centerline {}
\centerline {\bf {2. Quantization of pure gravity}}
\centerline {}

The problem of quantization of two-dimensional pure gravity
was studied in [1-4]. In [4,6] the author described an
anomaly-free quantization of a two-dimensional string whose
constraint system is the same as the constraint system of
two-dimensional pure gravity in the representation (1.16).
It gives us the possibility to apply here the methods
developed in [4,6].

Let \ $a=0, \, 1 $ \ and \ $ \eta_{ab}=diag (-1, \, 1) $.
In the gauge \ $u=1 \,
\ v=0 $ \ the Heisenberg equations for the fields \ $r^a \,, \
 \pi^a =\eta^{ab} \pi_b $ \ look like
$$
\left (\, \frac {\partial^2} {\partial t^2} -
\frac {\partial^2} {\partial \sigma^2} \, \right) \, r^a=0 \,, \ \ \
\left (\, \frac {\partial^2} {\partial t^2} -
\frac {\partial^2} {\partial \sigma^2} \, \right) \, \pi^a=0\,. \eqno (2.1)
$$
Therefore, the fields \ $r^a $ \ and \ $ \pi^a $ \ contain
both positive and negative-frequency modes:
$$
r^a (\sigma) = \frac {x^a} {\sqrt {4\pi}} +
\frac {i} {\sqrt {4\pi}} \, \sum_{n\neq 0} \, \frac {1} {n} \,
( \alpha^a_n\,e^{in\sigma}+\bar{\alpha}^a_n\,e^{-in\sigma}\,)\,,
$$
$$
\pi^a (\sigma)=\frac {p^a} {\sqrt {\pi}} +
\frac {1} {\sqrt {4\pi}} \, \sum_{n\neq 0} \,
(
\alpha^a_n\,e^{in\sigma}+\bar{\alpha}^a_n\,e^{-in\sigma}\,)\,.
  \eqno (2.2)
$$
We assume that \ $ \alpha^a_0\equiv\bar {\alpha} ^a_0\equiv p^a $.
 From the reality conditions for the fields (2.2) it follows that
$$
x^{a\dag} =x^a \,, \ \ \ \alpha^{a\dag}_n =\alpha^a_
{-n} \,, \ \ \
 \bar{\alpha}^{a\dag}_n=\bar{\alpha}^a_{-n}\,. \eqno (2.3)
$$
The commutation relations (1.17) are equivalent to the
following commutation relations of new variables:
$$
[
\alpha^a_m,\,\alpha^b_n\,]=[\bar{\alpha}^a_m,\,\bar{\alpha}^b_n\,]=
m \,\eta^{ab} \, \delta_{m+n} \,, \ \ \ [ x^a, \, p^b
\,] = i \,\eta^{ab}\,. \eqno (2.4)
$$
The set of operators (1.16) is equivalent to the two series of operators
$$
 L_n=\frac{1}{2}\,\int_0^{2\pi}\,d\sigma\, e^{-in\sigma}
\, ({\cal E}+{\cal P}) \,,
$$
$$
 \bar{L}_n=\frac{1}{2}\,\int_0^{2\pi}\,d\sigma\, e^
{in\sigma} \, ({\cal E} - {\cal P}) \,, \ \ \ n=0, \, \pm1,
\ldots
 \eqno (2.5)
$$
Using (2.2) we find:
$$
 L_n=\frac{1}{2}\,:\sum_m\,\alpha^a_{n-m}\,\alpha_{am}:\,, \
 \ \bar{L}_n=\frac{1}{2}\,:\sum_m\,\bar{\alpha}^a_{n-m}\,
\bar {\alpha}_{am}:\,. \eqno (2.6)
$$
The ordering of the operators in (2.6) is determined
according to the general conditions of quantization and
plays a main role. The purpose of quantization is searching
such space of physical states on which all the operators
(2.6) go to zero and in which there is a mathematically
correct and positive definite scalar product.

In the present article we adopt two approaches to the
quantization of the investigated system.

The first approach is well-known. It was formulated by Dirac
and is described by the following scheme. Let us assume that
\ $ \{\chi_n \, \}$ \ is a complete set of first-class
constraints. Then the physical states satisfy the conditions

$$
\chi_n \, | \ \rangle _ {PD} =0\,. \eqno (2.7)
$$
From (2.7) the consistency conditions of the theory follow:
$$
[ \chi_m, \, \chi_n \,] = c^l_{mn} \, \chi_l\,. \eqno
(2.8)
$$
In (2.8) the coefficients \ $c^l_{mn} $ \ may be operator
quantities and should be located to the left of the
constraints\ $ \chi_l $.

Under quantization (2.7)-(2.8) there is the following
difficulty (for further detail see [6]). It follows from the
conditions (2.7) that all the physical states do not depend
on certain initial dynamic variables. For this reason the
following problems arise:

 a) Determining the scalar product on the physical state
 space;

b) Calculating the matrix elements relative to the physical
states.

This is because not all the initial dynamic variables are
operators in the physical state space. Therefore the matrix
elements of these variables are not defined in physical
space. Though the observable values do not depend on the
indicated dynamic variables, nevertheless, serious
difficulties may arise when the matrix elements of the
observable quantities are calculated in physical space.

Further quantization (2.7)-(2.8) we shall name by the first
method of quantization.

In [6] a different method of quantization was applied to the
system (2.4),(2.6). The idea of this method of quantization
consists in some weakening of the Dirac conditions (2.7) by
replacing them with the conditions
$$
\langle \, P \, | \,\chi_m \, | \, P \,\rangle_G=0\,. \eqno (2.9)
$$
Here index \ $P $ \ numbers the physical states. In (2.9)
there is an averaging in all gauge degrees of freedom but
not in physical degrees of freedom.

The quantization conditions (2.9) are similar to the
Gupta-Bleuler conditions in electrodynamics when the
equality \ $ \partial _ {\mu} A ^ {\mu} =0 $ \ is only
satisfied in the sense of the mean value, and also to the
quantization conditions in the usual string theory when the
Virasoro algebra generators also only satisfy the conditions
\ $L_n=0 $ \ in the sense of the mean value. In this case,
averaging is performed relative to the physical states.

 The fundamental difference between the quantization method
 proposed here and the Gupta-Bleuler quantization and the
 generally accepted string quantization is that in our
 approach the complete state space has a positive definite
 scalar product. Below it is shown that this fact can be
 used to make an anomaly-free quantization of a
 two-dimensional string.

The conditions for consistency of the theory, which replace
the Dirac conditions (2.8), now have the form
$$
 \langle\,P\,|\,[\chi_m,\,\chi_n\,]\,|\,P\,\rangle_G=0\,.
 \eqno (2.10)
$$
The physical sense of the conditions (2.10) is as follows.
Let us assume that the Hamiltonian of the system has the
form used in the generally covariant theories: \ $ {\cal H}
_T =\sum \, v_m\chi_m $. We assume that at time \ $t $ \ the
conditions (2.9) are satisfied. At an infinitesimally close
time \ $t +\delta \, t $ \ the constraint \ $ \chi_n $ is
given by
$$
\chi_n (t +\delta t) = \chi_n (t) +i \,\delta t \,\sum_m \, v_m \,
 [\chi_m, \, \chi_n \,] (t)\,.
$$
 Therefore the self-consistency conditions (2.10) yield the
 equalities (2.9) at any time.

The quantization method (2.9)-(2.10) is subsequently called
as the second method of quantization.

We initially apply the first quantization method to the
model (2.6).

Let's introduce the notations
$$
x _ {\pm} =x^0\pm x^1 \,, \ \
 \alpha^{(\pm)}_m=\alpha^0_m\pm\alpha^1_m\,, \ \
 \bar{\alpha}^{(\pm)}_m=\bar{\alpha}^{(0)}_m\pm\bar{\alpha}^{(1)}_m\,.
   \eqno (2.11)
$$
The nonzero commutation relations of the new variables are obtained using (2.4):
$$
[ \alpha^{(+)}_m,\,\alpha^{(-)}_n\,]=
[\bar{\alpha}^{(+)}_m,\,\bar{\alpha}^{(-)}_n\,]=-2m\,\delta_{m+n}\,,
$$
$$
[
x_+,\,\alpha^{(-)}_n\,]=[x_-,\,\alpha^{(+)}_n\,]=-2i\,\delta_n\,.
   \eqno (2.12)
$$
Let's write operators (2.6) in the new variables:
$$
 L_n=-\frac{1}{2}\,\sum_m\,\alpha^{(+)}_{n-m}\,\alpha^{(-)}_m=
- \frac{1}{2}\,\sum_m\,\alpha^{(-)}_{n-m}\,\alpha^{(+)}_m\,.
  \eqno (2.13)
$$

Whenever possible subsequently, for the operators with a bar
we do not write those relations which are exactly the same
as those for the operators without a bare. By definition, in
(2.13) the ordering operation implies that either the
elements \ $ \alpha ^ {(+)} $ \ are placed to the left of
all the operators \ $ \alpha ^ {(-)} $, or the converse is
true. Both these orderings are equivalent (see [4,6]).

Let's make the canonical transformation
$$
\alpha\rightarrow U_M \,\alpha \, U_M ^ {\dag} \,, \ \
\bar {\alpha} \rightarrow U_M \,\bar {\alpha} \, U_M ^ {\dag} \,, \ \
x\rightarrow U_M \, x \, U_M ^ {\dag} \,,
$$
$$
 U_M=\exp\left(\frac{i\,M^2\,x_+}{2p_+}\,\right)\,.
$$
At this canonical transformation only the variables \ $
\alpha ^ {(-)} _ 0 $ \ and \ $x_- $ \ change, as given by
the formulas
$$
 \tilde{\alpha}^{(-)}_0=U_M\,\alpha^{(-)}_0\,U_M^{\dag}=
 \alpha^{(-)}_0-
\frac {M^2} {p _ +} \,,
$$
$$
\tilde {x} _ - =U_M \, x_- \, U_M ^ {\dag} =x_- +
M^2 \,\frac {x _ +} {p _ +} \,. \eqno (2.14)
$$
For monotony, in this section we introduce the notation \ $
\tilde {\alpha} ^ {(-)} _m =\alpha ^ {(-)} _ m $ \ for \
$m\neq 0 $.

Subsequently, instead of the operators (2.13) we use the
operators
$$
 L_n=-\frac{1}{2}\,\sum_m\,\alpha^{(+)}_{n-m}\,\tilde{\alpha}^{(-)}_m\,.
   \eqno (2.15)
$$
The reason for this substitution will become clear in the
following sections.

Let's define the vector state space in which the dynamic
variables of the system act, as linear operators. We present
the complete state space \ $H _ {CD} $ \ as the tensor
product of the gauge state space \ $H_G $\ and the physical
state space \ $H_{PD} $:

$$
H _ {CD} =H_G\otimes H_{PD}\,. \eqno (2.16)
$$
The space \ $H_G $\ is generated by its vacuum vector \ $ |
\, 0; \, G \,\rangle $, which is determined by the following
properties:
$$
\alpha^0 _ {-m} \, | \, 0; \, G \,\rangle=0 \,, \ \ \
\alpha^1_m \, | \, 0; \, G \,\rangle=0 \,, \ m > 0 \,, \ \ \
\langle \, 0; \, G \, | \, 0; \, G \,\rangle=1 \,. \eqno (2.17)
$$
The basis of the space \ $H_G $\ consists of vectors of the tipe
$$
 \alpha^0_m\ldots\bar{\alpha}^0_n\ldots\alpha^1_{-l}
 \ldots\bar{\alpha}^1_{-r}\,|\,0;\,G\,\rangle\,, \ \ \ m, \,
 n, \, l, \, r > 0\,. \eqno (2.18)
$$
Thus, \ $H_G $\ is a Fock space with a positive-definite
scalar product.

The basis in the physical Dirac state space \ $H _ {PD} $ \
consists of two series of states\ $ | \, k \,\rangle _ {D}
\,, \ k = (k^0, \, k^1 \,) $, possessing the following
properties:
$$
\tilde {\alpha} ^ {(-)} _m \, | \, k \,\rangle_D =
\tilde {\bar {\alpha}} ^ {(-)} _m \, | \, k \,\rangle_D=0 \,, \ \ \
m=0, \, \pm 1, \ldots \eqno (2.19)
$$
The relations (2.12) with \ $m=0 $ \ are rewritten in the
form
$$
( p^2_a+M^2 \,) \, | \, k \,\rangle_D=0\,. \eqno (2.20)
$$
From here it is seen that vectors of the basis set \ $ | \,
k \,\rangle _ {D} $ \ are split into two series of vectors \
$ | \, k\pm \,\rangle _ {D} $ \, each parametrized by a
single continuous real parameter. For example,
$$
p^1 \, | \, k\pm \,\rangle_{D} = \pm k \, | \, k\pm
\,\rangle_{D} \,, \ \ \ p^0 \, | \, k\pm \,\rangle_{D}=
\pm\sqrt {k^2+M^2} \, | \, k\pm \,\rangle_{D} \,, \ \ \
-\infty < k < + \infty\,. \eqno (2.21)
$$

Since the operators \ $p^a $ \ are Hermitian, the scalar
products
$$
 \langle\,k\pm\,|\,k'\pm\,\rangle_D=\delta(k-k')\,, \ \ \
\langle \, k- \, | \, k ' + \, \rangle_D=0 \eqno (2.22)
$$
are self-consistent.

Quantization conditions similar to (2.19) were used in
[1,2,4,6] and much earlier by Dirac in electrodynamics [8].

{ \bf {Note.}} { \it {We shall emphasize that the variables
\ $ \alpha^{a}_n $, \ $ \bar {\alpha}^{a}_n \,, \
n\neq 0 $, being linear operators in the space \ $H_G $, are
not generally operators in the space \ $H_{PD} $.
 Really, as a result of the action of the operators \ $
 \alpha ^ {(+)}_n $ \ and \ $ \bar {\alpha}^{(+)}_n $ \ on
 the vector \ $ | \, k\pm \,\rangle_D $ we obtain vectors
 which do not belong to the physical space \ $H_{PD} $.}} \\

Due to (2.13) and (2.19) the following equalities hold:
$$
 L_n\,|\,p\,\rangle_D=\bar{L}_n\,|\,p\,\rangle_D=0\,, \ \ \
 | \, p \,\rangle_D\in H_{PD}\,. \eqno (2.23)
$$
Thus, the model (2.6) is quantized using the first method.

Now we quantize this model using the second method. Let's
postulate that the complete state space \ $H_C $, in which
the initial variables act, is represented as the tensor
product
$$
H_C=H_G\otimes H_P\,. \eqno (2.24)
$$
Here the space \ $H_G $ \ is defined according to (2.17),
(2.18). The space \ $H_P $ \ has a basis with the properties
(2.20)-(2.22). If the vector is \ $ | \, p \,\rangle\in H _
{P} $, it satisfies the conditions (2.19) with \ $m=0 $.

We draw attention to that fact that the operators \ $ \alpha
^ {(+)} _ n $ \ and \ $ \bar {\alpha} ^ {(+)} _n $ \ with \
$n\neq 0 $ \ (or combinations of these) act in the space \
$H_G $, but their action is not determined on any one vector
from the space \ $H_P $. This distinguishes \ $H_P $ \ from
\ $H _ {PD} $ \ (see (2.19)). Thus, the complete state space
(2.24) is the tensor product of the spaces in which the
corresponding operators act. It is obvious that the space
(2.24) has a positive definite scalar product.

For further calculations it is necessary to determine the
ordering of the operators. The ordering (2.15) is equivalent
to the ordering
$$
L_0 =\frac {1} {2} \, ( p_a^2+M^2) -
\sum _ {m > 0} \, ( \alpha^0_m \,\alpha^0 _ {-m} -
\alpha^1 _ {-m} \, \alpha^1_m \,) \,, \eqno (2.25)
$$
which we shall use subsequently.

In our opinion, in this particular model the most convenient
physical states satisfying the conditions (2.9) are the
states which are coherent in gauge degrees of freedom. Let's
consider in space \ $H_G $ \ the coherent state
$$
 |\,z,\,\bar{z};\,G\,\rangle=\prod_{m>0}\,\exp\left\{-
 \frac{1}{2m}\,(|\,z^0_{-m}\,|^2+|\,z^1_m\,|^2
 +|\,\bar{z}^0_{-m}\,|^2+|\,\bar{z}^1_m\,|^2)+ \right.
$$
$$
\left. + \frac {1} {m} \, ( z^0_{-m} \, \alpha^0_m +
 z^1_m\,\alpha^1_{-m}+\bar{z}^0_{-m}\,\bar{\alpha}^0_m+
\bar {z} ^1_m \,\bar {\alpha} ^1_{-m}
\,) \, \right \} \, | \, 0; \, G \,\rangle\,. \eqno (2.26)
$$
Here \ $z^a_m $ \ and \ $ \bar {z} ^a_m $ \ are complex
numbers. Further let's assume that \ $z^a_0 =\bar {z} ^a_0 $
\ and \ $z ^ {a *} _ m=z^a _ {-m} $. The asterisk from above
means complex conjugation. By definition we have
$$
z^{(\pm)}_m=z^0_m\pm z^1_m , \ \ \bar{z}^{(\pm)}_m =
\bar {z}^0_m\pm \bar{z}^1_m\,.
$$
Let's also introduce the notation
$$
\tilde{z}^{(-)} _m = {z}^{(-)}_m \,, \ \
\tilde{\bar {z}} ^{(-)}_m =\bar {z}^{(-)}_m \,, \ \ m\neq 0\,,
$$
$$
 \tilde{z}^{(-)}_0=z^{(-)}_0-\frac{M^2}{z^{(+)}_0}\,.
 \eqno(2.27)
$$
Everywhere in this article it is assumed that \ $z^0_0 > 0$.

We denote the basis vectors in space \ $H_P $ \ as \ $ | \,
z_0\pm \,\rangle_P \,, \ z_0^a\equiv k^a $. We have:
$$
p^a \, | \, z_0\pm \,\rangle_P =\pm z^a_0 \, | \, z_0\pm
\,\rangle_P \,, \ \ \ \tilde {z} ^ {(-)} _0=0\,. \eqno (2.28)
$$
The last equality in (2.28) is a consequence of the relation
(2.19) with \ $m=0 $.

Let us assume that the sets of complex numbers \ $ \{z, \,
\bar {z} \, \} $ \ satisfy the equations (2.28) with the
upper sign and
$$
L_n (z) \equiv - \frac{1}{2}\,\sum_m\,z^{(+)}_{n-m}\,\tilde{z}^{(-)}_m
=0 \,, \ \
L_n (\bar {z}) =0 \,, \ \ n=0, \, \pm 1, \ldots
  \eqno (2.29)
$$
These equations are rewritten in the more convenient form if
we use the following notations:
$$ z^{(+)}(\sigma)\equiv\frac{1}{\sqrt{2\pi}}
\,\sum_n\,e^{in\sigma}\,z^{(+)}_n\,, \ \
 \bar{z}^{(+)}(\sigma)\equiv\frac{1}{\sqrt{2\pi}}\,
\sum_n \, e ^ {-in\sigma} \, \bar {z} ^ {(+)} _n\,,
$$
$$
 z^{(+)}(\sigma)\,\tilde{z}^{(-)}(\sigma)=0\,, \ \ \
 \bar{z}^{(+)}(\sigma)\,\tilde{\bar{z}}^{(-)}(\sigma)=0\,.
 \eqno (2.29 ')
$$
The functions \ $ z ^ {(+)} (\sigma), \ \tilde {z} ^ {(-)}
(\sigma) $, \ $ \bar {z} ^ {(+)} (\sigma) $ \ and \ $ \tilde
{\bar {z}} ^ {(-)} (\sigma) $ \ are real and periodic and
their zero harmonics satisfy (2.27) and (2.28).

The states
$$
| \, z, \, \bar {z} \pm \,\rangle_P\equiv | \, \pm z, \,
\pm\bar {z}; \, G \,\rangle
\otimes | \, z_0\pm \,\rangle_P \eqno (2.30)
$$
are called basic physical states if the conditions (2.28)
and (2. $ 29 ' $) are satisfied.

According to the given definitions the basic physical states
have the following properties:
$$
 \alpha^0_{-m}\,|\,z,\,\bar{z}\pm\,\rangle_P=
\pm z^0_{-m}\,|\,z,\,\bar{z}\pm\,\rangle_P\,,
$$
$$
 \alpha^1_{m}\,|\,z,\,\bar{z}\pm\,\rangle_P=
\pm z^1 _ {m} \, | \, z, \, \bar {z} \pm \,\rangle_P \,, \
 \ \ m\geq 0\,. \eqno (2.31)
$$

From the formulas (2.25), (2.29) and (2.31) it immediately
follows, that in our case the conditions (2.9) are
satisfied:
$$
 \langle\,z,\,\bar{z}\pm\,|\,L_n\,|\,z,\,\bar{z}\pm\,\rangle_P=0\,,
 \ \ \
 \langle\,z,\,\bar{z}\pm\,|\,\bar{L}_n\,|\,z,\,\bar{z}\pm\,\rangle_P=0\,.
    \eqno (2.32)
$$

Let's check whether the self-consistency conditions (2.10)
are satisfied. For this it is enough to be convinced that
$$
 \langle\,z,\,\bar{z}\pm\,|\,(\,L_nL_{-n}-
 L_{-n}L_n\,)\,|\,z,\,\bar{z}\pm\,\rangle_P=0\,. \eqno
 (2.33)
$$
The simple calculation shows that
$$
L_nL_{-n}
= \frac{1}{2}\,\sum^n_{m=1}\,m
(n-m\,)+n\,(\tilde {\alpha}^1_0\,)^2+
 2n\,\sum^{n}_{m=1}\,\tilde{\alpha}^1_{-m}\,\tilde{\alpha}^1_m+
$$
$$
 +\sum^{\infty}_{m=n+1}\,(n+m)\,\tilde{\alpha}^1_{-m}
\tilde {\alpha}^1_m+
\sum^{\infty}_{m=n+1}\,(m-n)\,\tilde{\alpha}^0_m\tilde{\alpha}^0_{-m}
+:L_nL _ {-n}: \,. \eqno (2.34)
$$
Similarly
$$
L_{-n}L_{n}
=\frac{1}{2}\,\sum^n_{m=1}\,m
(n-m\,)+n\,(\tilde{\alpha}^0_0\,)^2+
 2n\,\sum^{n}_{m=1}\,\tilde{\alpha}^0_{m}\,\tilde{\alpha}^0_{-m}+
$$
$$
 +\sum^{\infty}_{m=n+1}\,(n+m)\,\tilde{\alpha}^0_{m}
\tilde{\alpha}^0_{-m}+
\sum^{\infty}_{m=n+1}\,(m-n)\,\tilde{\alpha}^1_{-m}\tilde{\alpha}^1_{m}
+:L_{-n} L_{n}: \,. \eqno (2.35)
$$
Here the operators \ $ \tilde {\alpha} ^a $ \ are expressed
in terms of the operators \ $ {\alpha} ^ {(+)} \, \ \tilde
{\alpha} ^ {(-)} $ \ in the same way that the operators \ $
\alpha^a $ \ are expressed in terms of the operators $
{\alpha} ^ {(+)} \, \ {\alpha} ^ {(-)} $.

Since, \ $: \, L_nL _ {-n} \,:\equiv: \, L_{-n} L_{n} \,: $,
from the last two equalities we have
$$
L_nL _ {-n} -L _ {-n} L_n=2n \, L_0 \,. \eqno (2.36)
$$
The ordering on the right-hand side of equality (2.36) is
defined in accordance with (2.25). From (2.36) it can be
seen that the equations (2.33) are satisfied, that is, the
self-consistency conditions (2.10) are satisfied.

Note that generally
$$
 \langle\,z,\,\bar{z}\pm\,|\,L_n\,L_{-n}\,|\,z,\,
\bar {z} \pm \,\rangle_P\neq 0\,. \eqno (2.37)
$$

We see that the second method also results in a
self-consistent quantum theory of the model (2.6).

Shortly we shall discuss the superposition principle in the
second quantization method.

Let's assume that the states \ $ | \, z, \, \bar
{z} \pm \,\rangle_P $ \ and \ $ | \, z ', \,
\bar {z} '\pm \,\rangle_P $ \ are physical. Is the state
$$
 |\,z,\,\bar{z}\pm\,\rangle_P+|\,z',\,\bar{z}'\pm\,\rangle_P
 \eqno (2.38)
$$
physical?

In our view, {\it {the superposition principle need not
necessarily be extended to nonphysical gauge degrees of
freedom.}} Therefore, if in the more complex theories using
the second quantization method, the superposition principle
in the space \ $H_G $ \ will appear restricted, in our
opinion this does not invalidate the method. In physical
state space the superposition principle is fully obeyed.

\centerline {}
\centerline {}
\centerline {\bf {3. Inclusion of matter}}
\centerline {}
 It can be seen from the (anti)commutation relations
(1.17-18) that boson and fermion fields have the following
expansions in terms of modes (cf. (2.2)-(2.4)):
$$
f (\sigma) = \frac {x} {\sqrt {4\pi}} +
\frac {i} {\sqrt {4\pi}} \, \sum _ {n\neq 0} \, \frac {1} {n} \,
( \alpha_n\,e^{in\sigma}+\bar{\alpha}_n\,e^{-in\sigma}\,)\,,
$$
$$
\pi (\sigma) = \frac {p} {\sqrt {\pi}} +
\frac {1} {\sqrt {4\pi}} \, \sum _ {n\neq 0} \,
( \alpha_n\,e^{in\sigma}+\bar{\alpha}_n\,e^{-in\sigma}\,)\,,
  \eqno (3.1)
$$
$$
 \phi(\sigma)=\frac{1}{\sqrt{2\pi}}\,\sum_n\,\beta_n\,e^{in\sigma}\,,
 \ \ \chi (\sigma) =
 \frac{1}{\sqrt{2\pi}}\,\sum_n\,\bar{\beta}_n\,e^{-in\sigma}\,.
  \eqno (3.2)
$$
Further we suppose \ $p\equiv\alpha_0\equiv\bar {\alpha} _0
$. Since the fields (3.1) and (3.2) are real, we have:
$$
x ^ {\dag} =x \,, \ \alpha ^ {\dag} _n =\alpha _ {-n} \,
\ \ \ \bar {\alpha} ^ {\dag} _n =\bar {\alpha} _ {-n} \, \ \
\ \beta ^ {\dag} _n =\beta _ {-n} \, \ \ \ \bar {\beta} ^ {\dag} _n =
\bar {\beta} _ {-n}\,.
 \eqno (3.3)
$$

The (anti)commutation relations (1.17-18) are equivalent to
$$
[
\alpha_m,\,\alpha_n\,]=[\bar{\alpha}_m,\,\bar{\alpha}_n\,]=
m \,\delta _ {m+n} \,, \ \ \ [x, \, p \,] = i \,, \eqno
(3.4)
$$
$$
 \{\,\beta_m,\,\beta_n\,\}=\{\,\bar{\beta}_m,\,\bar{\beta}_n\,\}=
\delta _ {m+n}\,. \eqno (3.5)
$$
We write out only non-zero commutation relations. The
operators (2.6) we shall denote further by \ $L ^ {(0)} _n $
\ and \ $ \bar {L} ^ {(0)} _n $ \, respectively. Taking into
account the contribution of the material degrees of freedom,
we write out the Fourier components of(2.5):
$$
L_n = {L} ^ {(0)} _n +\frac {1} {2} \, \sum_m \,
\left [\alpha _ {n-m} \alpha_m +
 \left(m-\frac{n}{2}\,\right)\,\beta_{n-m}\beta_m\,\right] \eqno (3.6)
$$

 It is convenient to begin constructing the unitary
 transformation indicated at the end of the Introduction,
 which solves the quantization problem, by defining the
 creation and annihilation operators of the field (1.21). In
 other words, our first task is to construct both boson and
 fermion creation and annihilation operators of matter which
 commute with all the operators (3.6). We can see that the
 problem with slightly weaker conditions has a solution.
 This is sufficient for our purposes.

Let's consider the "gravitationally dressed" operators of
the material fields:
$$
A_m =\sum_n \, {\cal M} _ {m, n} \, \alpha_n \,, \ \
\bar {A} _m =\sum_n \,\bar {\cal M} _ {m, n} \, \bar {\alpha} _n \,, \eqno (3.7)
$$
$$
B_m =\sum_n \, {} ^F {\cal M} _ {m, n} \, \beta_n \,, \ \
\bar {B} _m =\sum_n \, {} ^F\bar {\cal M} _ {m, n} \, \bar {\beta} _n\,. \eqno (3.8)
$$
The infinite-dimensional matrices \ $ {\cal M}_{m, n} \, \
\bar {\cal M} _ {m, n} $, $ {} ^F {\cal M} _ {m, n} $ \ and
\ $ {} ^F\bar {\cal M} _ {m, n} $ \ in (3.7) and (3.8) are
determined in the Appendix. The elements of these matrixes
depend on the operators \ $ (x _ +/p _ + \,, \ \alpha ^
{(+)} _ m/p _ + $, $ \bar {\alpha} ^ {(+)} _m/p _ + \,) $,
whose relative commutators are equal to zero. Therefore all
the matrix elements in (3.7), (3.8) mutually commute.

It is easy to check by means of direct calculations that the
nonzero commutators of the operators (3.7) and (3.8) have
the following form:
$$
[ A_m, \, A_n \,] = [\bar {A} _m, \, \bar {A} _n \,] =
 m \,\delta _ {m+n} \,,
   \eqno (3.9a)
$$
$$
 \{\,B_m,\,B_n\,\}=\{\,\bar{B}_m,\,\bar{B}_n\,\}=\delta_{m+n}\,.
   \eqno (3.9b)
$$
Using (3.4),(3.7), and (A.14) we find:
$$
[ A_m, \, A_n \,] =
\sum_l \, l \, {\cal M} _ {m, l} \, {\cal M} _ {n, -l} =
-n \,\sum _ {l} \, {\cal M} _ {m, l} \, {\cal M} ^ {-1} _ {l, -n} =
m \,\delta _ {m+n}\,.
   \eqno (3.10)
$$
The equalities (3.9a) are thereby established. As a result
of (3.5), (3.8) and (A.$14 ' $) we have:
$$
\{\, B_m, \, B_n \, \} =\sum_l \, {} ^F {\cal M}_{m, l} {} ^F {\cal M}_{n, -l} =
\sum_l \, {} ^F {\cal M} _ {m, l} {} ^F {\cal M} ^ {-1} _ {l, -n}\,. \eqno(3.11)
$$
From this it follows that the commutation relations (3.9b) are valid.

The operators (3.7) and (3.8) introduced here differ only
negligibly from the DDF-operators used in the string theory
(see [9,10]).

From the given definitions it is easy to see that:
$$
[ \alpha^{(+)}_m,\,A_n\,]=[\alpha^{(+)}_m,\,\bar{A}_n\,]= [
\alpha^{(+)}_m,\,B_n\,]=[\alpha^{(+)}_m,\,\bar{B}_n\,]=0\,.
  \eqno (3.12)
$$
The relations (3.12) remain valid if instead of \ $ \alpha ^
{(+)} _ m $ \ we substitute $ \bar {\alpha} ^ {(+)} _m $ \
or \ $x _ + $.

Now, instead of $ \alpha ^ {(-)} _ m $ \ we must introduce
the variables $ \tilde {\alpha} ^ {(-)} _m $ \ into the
theory, which conserve the previous form of the commutation
relations with the variables $ \alpha ^ {(+)} _ m $ \ and
have zero commutators with the new variables (3.7), (3.8).

From the definition (3.7) we find:
$$
[ \alpha^{(-)}_m,\,A_n\,]=\sum_l\,[\alpha^{(-)}_m,\, { \cal
M} _ {n, l} \,] \, \alpha_l\,.
$$
We use (A.16) and also reverse equalities (3.7). As a
result, we obtain
$$
[ \alpha^{(-)}_m,\,A_n\,]=-\frac{2n}{p_+}\,\sum_p\,
\left (\, \sum_l \, {\cal M} _ {n, l-m} \,
 {\cal M} ^ {-1} _ {l, p} \, \right) \, A_p \,,
\ \ \ m\neq 0\,.
$$
Here the sum in brackets is calculated using formulas (A.3),
(A.10) and (A.12). Thus, we find
$$
[ \alpha^{(-)}_m,\,A_n\,]=-\frac{2n}{p_+}\,\sum_p\, { \cal
M} ^ {-1} _ {m, p-n} \, A_p \,, \ \ \ m\neq 0\,,
$$
$$
[ \alpha^{(-)}_0,\,A_n\,]=-\frac{n}{p_+}\,A_n\,. \eqno(3.13)
$$
Equality (3.11) here was taken into account. The following
relations also hold:
$$
[ \alpha^{(-)}_0,\,\bar{A}_n\,]=-\frac{n}{p_+}\,\bar{A}_n\,,
\ \ \ [ \alpha ^ {(-)} _ m, \, \bar {A} _n \,] = 0 \,, \ \ \
m\neq 0\,. \eqno(3.14)
$$

Similarly, using formulas (3.8), (A.4), (A.11), (A.12) and
(A.14) we obtain:
$$
[ \alpha^{(-)}_m,\,B_n\,]=-\frac{1}{p_+}\,\sum_p\,(n+p)\, {
\cal M} ^ {-1} _ {m, p-n} \, B_p \,, \ \ \ [ \alpha ^ {(-)}
_ m, \, \bar {B} _n \,] = 0 \,, \ \ \ m\neq 0 \,,
$$
$$
[ \alpha^{(-)}_0,\,{B}_n\,]=-\frac{n}{p_+}\,{B}_n\,, \ \ [
\alpha^{(-)}_0,\,\bar{B}_n\,]=-\frac{n}{p_+}\,\bar{B}_n\,.
     \eqno (3.15)
$$

It follows directly from formulas (3.13)-(3.15) and also
(3.9) and (3.12) that variables
$$
 \tilde{\alpha}^{(-)}_m=\alpha^{(-)}_m-\frac{1}{p_+}\,\sum_p\sum_q\,
{ \cal M} ^ {-1} _ {m, p+q} \cdot
 \left(A_p\,A_q-\frac{p-q}{2}\,B_p\,B_q\,\right)\,, \ \
 m\neq 0 \,,
$$
$$
 \tilde{\alpha}^{(-)}_0\equiv\tilde{\bar{\alpha}}^{(-)}_0=\alpha^{(-)}_0
- \frac{1}{2p_+}\,\sum_p\,\{\,A_p\,A_{-p}+\bar{A}_p\,\bar{A}_{-p}+
 p\,(B_{-p}\,B_{p}+\bar{B}_{-p}\,\bar{B}_{p}\,)+2M^2\,\} \eqno (3.16)
$$
commute with all the operators \ $A_n, \ \bar{A}_n, $ \ $B_n
$ and $\bar {B}_n $ \,:
$$
[ \tilde {\alpha} ^ {(-)} _m, \, A_n \,] =
[ \tilde {\alpha} ^ {(-)} _m, \, \bar {A} _n \,] =
[ \tilde {\alpha} ^ {(-)} _m, \, B_n \,] =
[ \tilde {\alpha} ^ {(-)} _m, \, \bar {B} _n \,] =
 0\,. \eqno (3.17)
$$
The number \ $M^2 $ \ in (3.16) can also be regarded as the
result of normal ordering (compare with (2.14)).

If in the formulas (3.13)-(3.17) all quantities without a
bar are replaced by the same quantities with a bar and
simultaneously all quantities with a bar are replaced by the
same quantities without a bar, these formulas remain valid.

Now we shall define the normal ordering of the creation and
annihilation operators (3.7) and (3.8). These operators are
by definition assumed to be normally ordered if all the
creation operators \ $A _ {- | n |}, \ \bar {A} _ {- |n |},
$ \ $B _ {- | n |}, \ \bar {B} _ {- |n |} $ are placed to
the left of all annihilation operators \ $A _ {| n |}, \
\bar {A} _ {|n |}, $ \ $B _ {| n |}, \ \bar {B} _ {|n |} $.

 The right-hand sides of the equalities (3.16) contain
 quadratic forms of the operators (3.7) and (3.8). These
 quadratic forms are represented as sums which are not
 normally ordered. However, it is easy to see, that the
 right-hand side of the equalities (3.16) can in fact be
 considered to be normally ordered, since the following
 equalities are satisfied
$$
\sum_p\sum_q \, {\cal M} ^ {-1} _ {m, p+q} \,
 \left(\,A_p\,A_q-\frac{p-q}{2}\,B_p\,B_q\,\right)
=:\sum_p\sum_q \, {\cal M} ^ {-1} _ {m, p+q} \,
 \left(\,A_p\,A_q-\frac{p-q}{2}\,B_p\,B_q\,\right):\,. \eqno
 (3.18)
$$
In order to prove the equalities (3.18) it is sufficient to establish that
$$
 \sum_p\,(A_p\,A_{-p}+p\,B_{-p}\,B_{p}\,)=
 \sum_p\,:(A_p\,A_{-p}+p\,B_{-p}\,B_{p}\,):=
$$
$$
=:\sum_p \, (A_p \, A _ {-p} +p \, B _ {-p} \, B _ {p} \,):\,. \eqno (3.19)
$$
The first equality in (3.19) follows directly from the
definition of orderings and the commutation relations (3.9).
In order to prove the second equality in (3.19), it is
necessary to take into account that formally \ $ \sum _
{n=1} ^ {\infty} \, n =\zeta (-1) $, where \ $ \zeta (s) $ \
is a Riemann zeta function:
$$
\zeta (s) = \sum _ {n=1} ^ {\infty} \, n ^ {-s} \eqno (3.20)
$$
The zeta function has a unique analytic continuation to the
point \ $s =-1 $ \ where \ $ \zeta (-1) = -1/12 $. This
regularization of the divergent sum \ $ \sum _ {n=1} ^
{\infty} \, n $ \ is now generally accepted. Therefore it is
possible to put that
$$
\left (\, \sum _ {n=1} ^ {\infty} \, n \,\right) -
 \left(\,\sum_{n=1}^{\infty}\,n\,\right)=\zeta(-1)-\zeta(-1)\,.
$$
Here the first divergent sum arises as a result of the
ordering of the boson operators and the second appears as a
result of the ordering of the fermion operators. From here
the second equality in (3.19) follows.

Thus, the right-hand sides of the equalities (3.16) can
equally well be taken to be normally ordered relative to the
operators (3.7) and (3.8) ore disordered and written in the
form of sums contained in (3.16). For some calculations the
disordered variant of the right-hand sides of (3.16) is more
convenient.

Now we shall prove the following commutation relations:
$$
[ \tilde{\alpha}^{(-)}_m,\,\tilde{\alpha}^{(-)}_n\,]= [
\tilde{\bar{\alpha}}^{(-)}_m,\,\tilde{\bar{\alpha}}^{(-)}_n\,]=
[
\tilde{\alpha}^{(-)}_m,\,\tilde{\bar{\alpha}}^{(-)}_n\,]=0\,.
   \eqno (3.21)
$$
Let \ $m\neq 0 $ \ and \ $n\neq 0 $. We take the  variables \
$ \tilde {\alpha} ^ {(-)} _m $ \ in the form (3.16). We then have
$$
[ \tilde{\alpha}^{(-)}_m,\,\tilde{\alpha}^{(-)}_n\,]=
[ \tilde{\alpha}^{(-)}_m,\,{\alpha}^{(-)}_n\,]
- \frac{1}{p_+}\,\sum_{p,q}\,[\tilde{\alpha}^{(-)}_m,
\, {\cal M} ^ {-1} _ {n, p+q} \,] \cdot
 \left(A_p\,A_q-\frac{p-q}{2}\,B_p\,B_q\,\right)\,.
 \eqno(3.22)
$$
Here we have used the definition (3.16) for \ $ \tilde
{\alpha} ^ {(-)} _n $ \ and the commutation relations
(3.17). On the right-hand side of (3.22) we replace \ $
\tilde {\alpha} ^ {(-)} _m $ \ by its value as given by
(3.16):
$$
[ \tilde{\alpha}^{(-)}_m,\,\tilde{\alpha}^{(-)}_n\,]=
 \frac{1}{p_+}\,\sum_{p,q}\,\{\,[\tilde{\alpha}^{(-)}_n,
\, {\cal M} ^ {-1} _ {m, p+q} \,] -
[ \tilde {\alpha} ^ {(-)} _m,
\, {\cal M} ^ {-1} _ {n, p+q} \,] \, \}
 \cdot\left(A_p\,A_q-\frac{p-q}{2}\,B_p\,B_q\,\right)+
$$
$$
+ \frac {1} {p _ +} \,\sum _ {p,q} \, {\cal M} ^ {-1} _ {m,
p+q} \,
\left\{\, [\alpha ^ {(-)} _ n, \, A_pA_q \,]-
 \frac{p-q}{2}\,[\alpha^{(-)}_n,\,B_pB_q\,]\,\right\}\,.
$$
Using formulas (A.17) and (3.13),(3.15), after redefining
the notation of the indices in some sums we transform this
last expression to give
$$
 \frac{1}{p^2_+}\,\left\{\,\sum_{p,q}\,\left[2m\,{\cal M} _
 {- (p+q), - (m+n)} \,
 \left(A_p\,A_q-\frac{p-q}{2}\,B_p\,B_q\,\right)
-2 \,\sum_r \, r \, {\cal M} ^ {-1} _ {m, q+r} \,
{ \cal M} ^ {-1} _ {n, p-r} \, A_p \, A_q + \right. \right.
$$
$$
\left. \left. + \frac{1}{2} \, \sum_r \,
(r+p) (r-q) \, {\cal M}^{-1}_{m, q+r} \, { \cal
M}^{-1}_{n,p-r}\,B_p\,B_q\,\right]\,\right\}-
\frac{1}{p^2_+} \ \{m\leftrightarrow n \,\}\,. \eqno(3.23)
$$
Here the parts of the sums over \ $r $ which are
antisymmetric in terms of the indices \ $m$ \ and \ $n$ are
obtained using the relations (A.18). As a result, all the
terms in (3.23) are mutually reduced. Thus, we have proven
that the commutator (3.22) is equal to zero. The relations
$$
[
\tilde{\bar{\alpha}}^{(-)}_m,\,\tilde{\bar{\alpha}}^{(-)}_n\,]=0\,,
\ \ \ m\neq 0 \,, \ \ n\neq 0
$$
are proved by exactly repeating these procedure.

Similarly, it is established that
$$
[ \tilde{\alpha}^{(-)}_0,\,\tilde{\alpha}^{(-)}_m\,]= [
\tilde{\alpha}^{(-)}_0,\,\tilde{\bar{\alpha}}^{(-)}_m\,]=0\,.
$$
The equalities
$$
[
\tilde{\alpha}^{(-)}_m,\,\tilde{\bar{\alpha}}^{(-)}_n\,]=0\,,
\ \ \ m\neq 0 \,, \ \ n\neq 0
$$
follow trivially from the fundamental commutation
relations (3.4), (3.5) and the definitions (3.16). The
validity of the commutation relations (3.21) is thus
completely proven.

From the definitions (3.16) and the commutation relations (3.12) we have also:
$$
[ \alpha^{(+)}_m,\,\tilde{\alpha}^{(-)}_n\,]=
[ \bar{\alpha}^{(+)}_m,\,\tilde{\bar{\alpha}}^{(-)}_n\,]=-2m\,\delta_{m+n}\,,
 \ \ \ [x _ +, \,\tilde {\alpha} ^ {(-)} _n \,] = [
\tilde{x}_-,\,\tilde{\alpha}^{(+)}_n\,]=-2i\,\delta_n\,. \eqno
(3.24)
$$
The explicit form of the new variable \ $ \tilde {x} _ - $
\ is not given here since this variable is not used below.

The initial variables \ $ \{x _ {\pm}, \, \alpha ^ {(\pm)}
_m, \, $
$\bar{\alpha}^{(\pm)}_m,\,\alpha_m,\,\bar{\alpha}_m,\,$ $
\beta_m, \, \bar {\beta} _m \, \} $ \ (or more accurately
their linear combination) are canonical one. It follows from
the commutation relations (3.9),(3.12), (3.17) and (3.21)
that the set of variables
$$
 \{x_{+},\,\tilde{x}_-,\,\alpha^{(+)}_m,\,
 \tilde{\alpha}^{-}_m,\,\bar{\alpha}^{(+)}_m,
\, \tilde {\bar {\alpha}} ^ {(-)} _m, \,
A_m, \, \bar {A} _m, \, B_m, \, \bar {B} _m \} \eqno (3.25)
$$
is also canonical.

Now we can determine the unitary transformation appearing in
(1.21). Let's define the unitary operator \ $U $ \ using the
following equalities:
$$
U \, x _ + = x _ + \, U \,, \ \ \ U \, x_- =\tilde {x} _ - \, U \,, \ \ \
U \,\alpha ^ {(+)} _ m =\alpha ^ {(+)} _ m \, U \,, \ \ \
 U\,\alpha^{(-)}_m=\tilde{\alpha}^{(-)}_m\,U\,,
$$
$$
U \,\alpha_m=A_m \, U \,, \ \ \ U \,\beta_m=B_m \, U \,, \ \ \
 U\,\bar{\alpha}^{(+)}_m=\bar{\alpha}^{(+)}_m\,U\, \,,
 \ldots ,\eqno (3.26)
$$
and so on for the remaining operators with a bar. It is
known, that equalities of the type (3.26) uniquely determine
the linear operator \ $U $ \ and this operator is unitary
[11].

We must express the operators (3.6) in terms of the new
variables. To do this we represent the operators \ $L ^
{(0)} _n $ \ from (3.6) in the form (2.13) and express the
operators \ $ \alpha ^ {(-)} _ m, \ \alpha_m $\ and $\beta_m
$ \ in terms of the operators \ $x _ +, \ \alpha ^ {(+)} _
m, \ \tilde {\alpha} ^ {(-)} _m, $ \ $A_m, \ B_m $ \ using
formulas (3.7), (3.8) and (3.16). As a result of simple
calculations using the sum rules (A.20) and (A.21) we come
to the following formulas:
$$
 L_n=-\frac{1}{2}\,\sum_m\,\alpha^{(+)}_{n-m}\,\tilde{\alpha}^{(-)}_m+
\frac {1} {4p _ +} \,\alpha ^ {(+)} _ n \, {\cal N} \,,
$$
$$
\bar {L} _n =-\frac {1} {2} \,
 \sum_m\,\bar{\alpha}^{(+)}_{n-m}\,\tilde{\bar{\alpha}}^{(-)}_m-
 \frac{1}{4p_+}\,\bar{\alpha}^{(+)}_n\,{\cal N} \,,
$$
$$
{ \cal N}=\sum_l\,\{A_{-l}\,A_l-\bar{A}_{-l}\,\bar{A}_l+
 l\,(B_{-l}\,B_l-\bar{B}_{-l}\,\bar{B}_l)\,\}\,. \eqno
 (3.27)
$$
For the same reason as in (3.19) the equality
$$
{ \cal N} =: {\cal N}: \eqno (3.28)
$$
takes place.

It is possible to be convinced of the validity of the
equalities (3.27) by a more simple way. For this it is
necessary to calculate the commutators of the operators
(3.7), (3.8), (3.16) with the operators \ $L_n $ \ in the
representation (3.6) and (3.27). The results of these
calculations coincide.

Using (3.26) and (3.27), we find that
$$
L_n=U \,\left\{L ^ {(0)} _n +
\frac {1} {2p _ +} \,\alpha ^ {(+)} _ n \, {\cal N} _0 \,\right\} \, U ^ {\dag}
\,,
$$
$$
\bar {L} _n=U \,\left\{\bar {L} ^ {(0)} _n-
 \frac{1}{2p_+}\,\bar{\alpha}^{(+)}_n\,{\cal N}_0 \,\right\}
 \, U ^ {\dag} \,,
$$
$$
{ \cal N} _0 =\sum _ {l > 0} \, [\alpha _ {-l} \alpha_l-
 \bar{\alpha}_{-l}\bar{\alpha}_l+l\,(\beta_{-l}\,\beta_l-
\bar {\beta} _ {-l} \, \bar {\beta} _l \,) \,]\,. \eqno (3.29)
$$
The relations (3.29) are an exact variant of (1.20). Though
the formulas (3.29) have a little more complex form than
(1.20), nevertheless there is a possibility of essential
further promoting on the path which was indicated in
Introduction.

We note that the operators (3.7),(3.8) used here differ
essentially from gravitationally dressed operators in [2].
For the operators (3.7), (3.8) we have:
$$
[ A_m,\,L_n\,]=\frac{m}{2p_+}\,\alpha^{(+)}_n\,A_m \eqno (3.30)
$$
and so on. We stress that it is impossible to construct a
set of operators which are expressed linearly in terms of
material field operators and also commuting with all
operators \ $L_n $ \ and \ $ \bar {L} _n $. (A similar
situation is encountered in theory of closed string when the
transverse degrees of freedom are described by
DDF-operators.) A different approach to the investigated
model was applied in [2]: the authors introduced new
operators \ $L ^ {\prime} _n $ \ and \ $ \bar {L} ^ {\prime}
_n $, which differ from the operators \ $L_n $ \ and \ $
\bar {L} _n $ by values proportional to Planck constant. The
operators \ $L ^ {\prime} _n $ \ and \ $ \bar {L} ^ {\prime}
_n $ have the same algebra as \ $L_n $ \ and \ $ \bar {L} _n
$. At the same time there is a complete set of
gravitationally dressed operators \ $ \{C_n, \, \bar {C} _n
\, \} $, describing the degrees of freedom of the material
fields and expressed linearly in terms of material field
operators, and also commuting with all operators \ $L ^
{\prime} _n $ \ and \ $ \bar {L} ^ {\prime} _n $. This fact
simplifies the formal quantization procedure. However,
essential difficulties are encountered when we attempt to
express the initial dynamic variables in terms of those
operators used for quantization.

In the present paper, unlike [2], we give explicit formulas
expressing the initial variables in terms of the new
variables (see (3.16)). It enables the calculations of
matrix elements of the metric tensor (1.2) (see Section 5).

\centerline {}
\centerline {}
\centerline {\bf {4. Physical state space}}
\centerline {}

The formulas obtained in Sections 2 and 3 can be used to
quantize the investigated model.

As a result of the existence of a unitary transformation
having the properties (3.26) and (3.29), we can confirm that
the state space of the system (3.6) is isomorphic to the
state space of the two noninteracting systems: pure gravity
and free fields (3.1), (3.2). Bearing in mind this unitary
transformation, we shall construct the physical state space
directly in the theory with interaction.

We first apply the first quantization method.

Let's define two families of states using the formulas (cf.
(2.19)):
$$
 \tilde{\alpha}^{(-)}_m\,|\,k\pm\,\rangle_D=
 \tilde{\bar{\alpha}}^{(-)}_m\,|\,k\pm\,\rangle_D=0\,, \ \
m=0, \, \pm1, \ldots , \eqno (4.1)
$$
$$
A_n \, | \, k\pm \,\rangle_D = \bar {A} _n \, | \, k\pm \,\rangle_D =
 B_n\,|\,k\pm\,\rangle_D=\bar{B}_n\,|\,k\pm\,\rangle_D=0\,,
 \ n > 0\,.
   \eqno (4.2)
$$
Let's impose also the constraints
$$
 A_0\,|\,k\pm\,\rangle_D=\bar{A}_0\,|\,k\pm\,\rangle_D=0\,,
$$
which are not necessary and are imposed only for
simplification of the formulas. In addition, we assume that
relations (2.21) and (2.22) are satisfied.

The reason why the quantity \ $M^2 > 0 $ was introduced into
the formulas (2.14) and (3.16) now becomes clear: as a
result of this introduction and condition (4.1) with \ $m=0
$ \ the operator \ $p _ + $ \ has no zero eigenvalues in the
physical space \ $H _ {PD} $. Therefore the unitary
transformation (3.26) is defined correctly and the operators
\ $A_n $ \ and so on can act on the states \ $ | \, k\pm
\,\rangle_D $.

All the physical states are linear combinations of basic
states having the form
$$
 |\,k\pm;\,n_i,\,\bar{n}_i,\,m_i,\,\bar{m}_i\,\rangle_D=
( A _ {-n_1} \ldots \bar {A} _ {-\bar {n} _1} \ldots
B _ {-m_1} \ldots \bar {B} _ {-\bar {m} _1} \ldots) \ | \, k\pm \,\rangle_D
\in H ^ {(\pm)} _ {PD} \,,
$$
$$
n_i, \, \bar {n} _i, \, m_i, \, \bar {m} _i > 0 \,, \eqno (4.3)
$$
$$
\left (\, \sum_i \, n_i +\sum_i \, m_i-
 \sum_i\,\bar{n}_i-\sum_i\,\bar{m}_i\,\right)=0\,. \eqno
 (4.4)
$$
The complete physical state space is represented as a direct
sum
$$
H _ {PD} =H ^ {(+)} _ {PD} \oplus H ^ {(-)} _ {PD}\,. \eqno
(4.5)
$$
As a result of relations (4.1),(4.4) and (3.9), we have:
$$
{ \cal N} \, | \ \rangle_P=0 \,, \qquad | \ \rangle_P\in H _
{PD}\,. \eqno (4.6)
$$
Now using (3.27),(4.1),(4.2) and (4.6) we obtain
$$
L_n \, | \ \rangle_P = \bar {L} _n \, | \ \rangle_P=0\,.
\eqno (4.7)
$$

It follows from the commutation relations (3.9) and
equalities (4.2), (2.22) that the scalar product in the
spaces \ $ H ^ {(+)} _ {PD} $ \ and \ $ H ^ {(-)} _ {PD} $ \
is positive definite, and also these spaces are mutually
orthogonal.

This implies that an anomaly-free quantization of the system
(3.6) has been performed using the first method.

Now we apply the second method of quantization.

By definition, the state space is generated by two series of
states \ $ | \, z, \, \bar {z} \pm \,\rangle_P $ \ having
the following properties. The states \ $ | \, z, \, \bar {z}
\pm \,\rangle_P $ \ satisfy the equations (2.28),(4.2) and
also (cf. (2.31)) the equalities
$$
 \frac{1}{2}\,(\alpha^{(+)}_{-m}+\tilde{\alpha}^{(-)}_{-m}\,)\,
| \, z, \, \bar {z} \pm \,\rangle_P =
\pm z^0_{-m}\,|\,z,\,\bar{z}\pm\,\rangle_P\,,
$$
$$
 \frac{1}{2}\,(\alpha^{(+)}_{m}-\tilde{\alpha}^{(-)}_{m}\,)\,
| \, z, \, \bar {z} \pm \,\rangle_P =
\pm z^1 _ {m} \, | \, z, \, \bar {z} \pm \,\rangle_P \,, \ \ m > 0 \,, \eqno (4.8)
$$
and similarly for the quantities with a bar.

From (2.28) it follows that
$$
 \tilde{\alpha}^{(-)}_0\,|\,z,\,\bar{z}\pm\,\rangle_P=0\,.
 \eqno (4.9)
$$
The basic states in he physical space are denoted by \ $ |
\, z, \, \bar {z} \pm; $
$\,n_i,\,\bar{n}_i,\,m_i,\,\bar{m}_i\,\rangle_P$. They are
constructed using the operators \ $A _ {-m}, \ldots, \ m > 0
$ \ in accordance with (4.3) and (4.4). The physical state
space \ $H_P $ \ is decomposed in the direct sum of the
orthogonal subspaces \ $H_P ^ {(+)} $ \ and \ $H_P ^ {(-)}
$. The scalar product in the space \ $H_P $ \ is positive
definite.

As a result of the commutation relations (3.12) and (3.17),
equations (4.8) and (4.9) remain valid also for physical
states if the physical states are "pure" relative to gauge
degree of freedoms. Here we understand under the "purity"
that all physical states have the same set of parameters \ $
\{z^a_m, \, \bar {z} ^a_m \, \}$.

We shall not demonstrate in detail that in this case
quantization conditions (2.9) and (2.10) are satisfied. It
follows directly from all above-stated. Here we merely stress
the following important fact: {\it {Averaging in (2.9) and
(2.10) needs to be spent only in space of gauge degree of
freedoms \ $H_G $ \ (see (2.17),(2.18)). Averaging in (2.9)
and (2.10) need not be performed over variables described by
the operators \ $A_m, \ \bar {A} _m, $ \ $B_m, \ \bar {B} _m
$.}}

\centerline {}
\centerline {}
\centerline {\bf {5. Calculation of averages}}
\centerline {}

 In the investigated model the most interesting quantity is
 the average value of the metric tensor (1.2) relative to
 the physical states. For this purpose, in accordance with
 (1.2), it is necessary to calculate the average of the
 expression \ $ \exp (2\rho) $ \ as the parameters \ $u $ \
 and \ $v $ \ are the numerical Lagrange multiplies.

 To begin calculations, suppose that the formulas (1.15)
valid in classical theory also hold in quantum theory. This
assumption allows us to express the unknown quantity in
terms of the quantum fields \ $ \pi^a, \ r^a $ \ (2.2) as
follows:
$$
\frac {\lambda} {\pi \, G} \, e ^ {2\rho} =
\pi_a\pi^a-r ' _a {r^a} ^ {\prime} +2 \,
( \pi^0\,{r^1}^{\prime}-\pi^1\,{r^0}^{\prime}\,)\,.
$$
With the help of (2.2) this equality is rewritten in the
convenient form for us:
$$
 e^{2\rho(\sigma)}=-\frac{G}{\lambda}\,\left(\,\sum_m\,
\alpha ^ {(+)} _ m \, e ^ {im\sigma} \, \right)
 \,\left(\,\sum_n\,\bar{\alpha}^{(-)}_n\,e^{-in\sigma}\,\right)\,.
 \eqno (5.1)
$$
In (5.1) the variables \ $ \bar {\alpha} _n ^ {(-)} $ \
should be expressed in terms of the new variables \ $ \bar
{\alpha} _n ^ {(+)} $, \ \ $ \tilde {\bar {\alpha}} _n ^
{(-)} $ \ and \ $ \bar {A} _m, \ \bar {B} _m, $ \ $A_m, \
B_m $ \ in accordance with (3.16). Then we can calculate the
average values of the expression (5.1) using the results of
the previous Section.

On the right-hand side of (5.1) all the variables inside the
first brackets commute with all the variables inside the
second brackets. Taking into account this fact and also the
results of Section 4, we can confirm that the following
formula is valid to calculate the averages relative to the
basis vectors:
$$
 \langle\,e^{2\,\rho}\,\rangle=-\frac{G}{\lambda}\,
 \langle\,\sum_m\,\alpha^{(+)}_m\,e^{im\sigma}\,\rangle\cdot
 \langle\,\sum_n\,\bar{\alpha}^{(-)}_n\,e^{-in\sigma}\,\rangle\,.
  \eqno (5.2)
$$

For calculation of averages in (5.2) the second method of
quantization is used.

We first calculate the average relative to the ground state
 \ $ | \, z, \, \bar {z} + \, \rangle_P $.

Let's express the variables \ $ \alpha ^ {(+)} _ m $ \ in the form
$$
\alpha ^ {(+)} _ m=a^0_m+a^1_m \,, \ \
 a^0_m=\frac{1}{2}\,(\alpha^{(+)}_m+\tilde{\alpha}^{(-)}_m\,)\,,
 \ \
 a^1_m=\frac{1}{2}\,(\alpha^{(+)}_m-\tilde{\alpha}^{(-)}_m\,)\,.
   \eqno (5.3)
$$
and use the formulas (4.8) and (2.28). Thus we obtain
$$
\langle \, z, \, \bar {z} +\,|\,\sum_m\,\alpha^{(+)}_m\,e^{im\sigma}\,|
\, z, \, \bar {z} + \, \rangle_P =
\sum_m \, z ^ {(+)} _ m \, e ^ {im\sigma}\,. \eqno (5.4)
$$
Similarly we find
$$
\langle \, z, \, \bar {z} +\,|\,\sum_n\,\tilde{\bar{\alpha}}^{(-)}_n\,
e^{-in\sigma} \, |
\, z, \, \bar {z} + \, \rangle_P =
 \sum_n\,\tilde{\bar{z}}^{(-)}_n\,e^{-in\sigma}\,. \eqno
 (5.5)
$$
It follows from (3.16),(3.18) and (4.2) that
$$
\langle \, z, \, \bar {z} + \, | \, ({\bar {\alpha}}^{(-)}_n-
\tilde {\bar {\alpha}}^{(-)}_n) \, |
\, z, \, \bar {z} + \, \rangle_P=0\,. \eqno (5.6)
$$

Combining formulas (5.2),(5.4),(5.5),(5.6) and (2. $ 29 '
$), we obtain:
$$
\langle \, z, \, \bar {z} + \, | \,
e ^ {2\rho (\sigma)} \, |
\, z, \, \bar {z}
      + \, \rangle_P =
- \frac{2\pi\,G}{\lambda}\,z^{(+)}(\sigma)\,
\tilde {\bar {z}} ^ {(-)} (\sigma) \eqno (5.7)
$$

To find more complex matrix elements it is necessary to make
additional calculations.

As a result of (A.2), we have
$$
 u^l=z^l\,\exp\left(il\,\frac{x_+}{2p_+}\,\right)\,
\exp (\xi _ ++\xi_-) \,,
$$
$$
\xi_+=\frac{l}{p_+}\,\sum_{m>0}\,\frac{1}{m}\,(-z^{-m}\,a^0_m+
z^m \, a^1 _ {-m} \,)\,, \ \
\xi_-=\frac{l}{p_+}\,\sum_{m>0}\,\frac{1}{m}\,(z^{m}\,a^0_{-m}-
z^{-m} \, a^1_{m} \,)\,. \eqno(5.8)
$$
Here \ $a^0_m $ \ and \ $a^1_m $ \ are defined according to
(5.3). As a result of the relations (4.8), when calculating
the matrix elements we need to order the operators in (5.8)
so that the functions of the value \ $ \xi _ + $ \ are to
the left of the functions of the value \ $ \xi_- $. It is
easy for achieving since the commutator
$$
\frac {1} {2} \, [ \xi _ +, \,\xi_- \,] =
 \left(\frac{l}{p_+}\,\right)^2\,\sum_{m>0}\,\frac{1}{m} \eqno (5.9)
$$
is a c-number commuting with \ $ \xi _ + $ \ and \ $ \xi_-
$. Then using the well-known Baker-Hausdorff formula we
obtain
$$
 u^l=\left\{\,\exp\left[-\left(\frac{l}{p_+}\,\right)^2\,
 \sum_{m>0}\,\frac{1}{m}\,\right]\,\right\}\,
 z^l\,\exp\left(il\,\frac{x_+}{2p_+}\,\right)\cdot e ^ {\xi
_ +} \, e ^ {\xi_-}\,. \eqno (5.10)
$$
Now the averages of expression (5.10) are calculated using
relations (4.8):
$$
\langle \, z, \, \bar {z} + \, | \,
u^l \, |
\, z, \, \bar {z} + \, \rangle_P\sim
 \left\{\,\exp\left[-\left(\frac{l}{p_+}\,\right)^2\,\sum_{m>0}\,
\frac {1} {m} \, \right] \, \right \}\cdot
\exp\left (-\frac {l} {p _ +}\sum _ {n\neq 0} \, \frac {z ^ {(+)} _ n} {n} \,
z ^ {-n} \, \right)\,. \eqno (5.11)
$$
Since the argument of the first exponential function on the
right-hand side of (5.11) is an infinitely large negative
number, the average of (5.11) is equal to zero. It means
that in calculations of the matrix elements under the
average sign the elements of the matrix \ $ {\cal M} ^ {-1}
_ {n, \, l} $ \ are in fact only nonzero for \ $l=0 $. The
indicated fact which applies equally to the elements of the
matrix \ $ \bar {\cal M} ^ {-1} _ {n, \, l} $, essentially
simplifies the calculations. Using (A.9) and (A.10) we
obtain:
$$
\bar {\cal M}^{-1}_{n,\,0}=\frac{\bar{\alpha}^{(+)}_n}{p_+}\,. \eqno
(5.12)
$$

Bearing this observation in mind and also formula (A.12) we
note that the matrix elements should be calculated using the
following relations:
$$
 \bar{\alpha}^{(-)}_m=\tilde{\bar{\alpha}}^{(-)}_m+
 \frac{\bar{\alpha}^{(+)}_m}{p^2_+}\,:\sum_p\, (
\bar{A}_{-p}\,\bar{A}_p+p\,\bar{B}_{-p}\,\bar{B}_p):\,, \ \
m\neq 0 \,, \eqno (5.13)
$$
and also the second formula (3.16) (for $m=0 $). Using
(5.2),(5.4), and (5.13), we can calculate the diagonal
matrix element of the metric tensor relative to the basis
state like (4.3), (4.4) \ $ | \, p \,\rangle = $
$|\,z,\bar{z}+;\,n_i,\,\bar{n}_i,\,m_i,\,\bar{m}_i\,\rangle$:
$$
 \langle\,p\,|\,e^{2\rho(\sigma)}\,|\,p\,\rangle=
- \frac{2\pi\,G}{\lambda}\,z^{(+)}(\sigma)\,
\left\{\tilde {\bar {z}} ^ {(-)} (\sigma) +
 \frac{{\bar{z}}^{(+)}(\sigma)}{(z^{(+)}_0)^2}\, ( \, \sum_i
\, n_i +\sum_i \, m_i \,) \,\right \}\,. \eqno (5.14)
$$
The functions \ $ {z} ^ {(+)} (\sigma) $, \ $ \bar {z} ^
{(+)} (\sigma) $ \ and \ $ \tilde {\bar {z}} ^ {(-)}
(\sigma) $ \ satisfy (2. $ 29 ' $). To derive formula (5.14)
we used equality (4.4) and assumed that the basis vector is
normalized.

Further calculation of averages, their study and
interpretation are outside the scope of the present work.

The calculation of matrix elements of the metric tensor
using the first method encounters serious difficulties.
Really, the variables \ $ \alpha ^ {(+)} _ m $ \ are not
operators in the physical state space \ $H _ {PD} $.
 Therefore averages having the form
$$
 \langle\,\sum_m\,\alpha^{(+)}_m\,e^{im\,\sigma}\,\rangle_{PD}\,,
 \ \ \
 \langle\,\sum_m\,\bar{\alpha}^{(+)}_m\,e^{-im\,\sigma}\,\rangle_{PD}
   \eqno (5.15)
$$
are not determined. As a result of the quantization
conditions (4.1) and (4.2) it would be possible to accept
that the average (5.2) is equal to zero if the matrix
element is calculated relative to generating states \ $ | \,
k\pm \,\rangle_D $. However, it does not rescue a general
situation since at calculating the matrix elements relative
to the excited states it is necessary to calculate the
averages from the expressions of the tipe (5.15) (see (5.12)
and (5.14)).

\centerline {}
\centerline {}
\centerline {\bf {6. Conclusions}}
\centerline {}

 In the present paper we have applied two methods of
 quantization to the theory of two-dimensional gravity.
 The first method using the Dirac prescription allows us to
 achieve complete quantization. It implies

 {\bf {a)}} constructing a physical state space with a
 positive definite scalar product;

 {\bf {b)}} the obvious expressing physically meaningful
 quantities in terms of those operators used to construct
 the physical state space.

 However the averages of the metric tensor cannot be
 calculated using the first method. This statement is fair
 at least in that case when the space of physical states is
 constructed using the operators, \ $A_n, \ B_n, \ldots, $ \
 determined in Section 3.

In addition to the problems (a) and (b) the second method of
quantization can also solve the following problems:

 {\bf {c)}} calculating the averages of the metric tensor
 relative to the physical states.

On the basis of these results we can conclude that the
second method of quantization should be used subsequently to
study other models.

To conclude we make the following remark.

The model (3.6) can easily be quantized in a "light cone"
gauge. In the terms used in the present study using this
gauge implies imposing the second-class constraints:
$$
\alpha ^ {(+)} _ m=0 \,, \ \ \tilde {\alpha} ^ {(-)} _m=0 \,, \ \
\bar {\alpha} ^ {(+)} _m=0 \,, \ \ \tilde {\bar {\alpha}} ^ {(-)} _m=0 \,, \ \
 m\neq 0\,. \eqno (6.1)
$$
 Thus in accordance with (5.2) and (5.6), the average of the
 metric tensor relative to the ground state is zero. On the
 other hand, in the second method of quantization, in
 accordance with (5.7) and (2.$29'$), the average of the
 metric tensor relative to the ground state is generally
 nonzero. From here we see that if the quantization method
 based on obvious resolution of the first-class constraints
 by imposing the special gauges is equivalent to the first
 method of quantization, then it is not equivalent to the
 second method of quantization. Under the second
 quantization method the structure the physical state space
 appears essentially richer than under the first
 quantization method (ore under the obvious resolution of
 the first-class constraints).

\centerline {}
\centerline {}
\centerline {\sf {Appendix}}
\centerline {}

 Let \ $ \tau $ \ be some "time-like" parameter and \ $z=e ^
{i\tau} $. We introduce the following operator functions:
$$
q (\tau) \equiv q (z) = \frac {x _ +} {2p _ +} +\frac {1} {i} \, \ln
z +\frac {i} {p _ +} \,\sum _ {n\ne 0} \, \frac {1} {n} \,
 \alpha ^ {(+)} _ n \, z ^ {-n}\,,
$$
$$
\bar {q} (\tau) \equiv \bar{q}(z)=\frac{x_+}{2p_+}+\frac{1}{i}\,\ln
z +\frac {i} {p _ +} \,\sum _ {n\ne
 0}\,\frac{1}{n}\,\bar{\alpha}^{(+)}_n\,z^{-n}\,. \eqno
 (A.1)
$$
By definition we have
$$
u (z) \equiv e^{i\,q(z)}=z\,\exp\left(i\frac{x_+}{2p_+}\right)\cdot
\exp\left (-\sum _ {n\ne
 0}\frac{\alpha^{(+)}_n}{np_+}\,z^{-n}\,\right)\,, \ \
\bar {u} (z) \equiv e ^ {i \,\bar {q} (z)}\,. \eqno (A.2)
$$

Let the contour \ $C $ \ in the \ $z $-plane go
counterclockwise once around the point $z=0$ . Let's define
four infinite-dimensional matrixes according to the
formulas:
$$
{ \cal M} _ {m, n} = \frac {1} {2\pi i} \, \oint_C
 \,\frac {dz} {z} \, z ^ {-n} \, u^m\,,
  \eqno (A.3)
$$
$$
{} ^F {\cal M} _ {m, n} =
\frac {1} {2\pi i} \, \oint_C \,\frac {dz} {z} \, z ^ {-n} \, u^m
\, \sqrt {\dot {q}}\,. \eqno (A.4)
$$
The definitions of the matrixes \ $ \bar {\cal M} _ {m, n} $
\ and \ $ {} ^F\bar {\cal M} _ {m, n} $ \ are obtained from
(A.3) and (A.4) by substituting \ $q\rightarrow\bar {q} $, \
and\ $u\rightarrow\bar {u} $. Hereinafter we have:
$$
 \dot{q}=\frac{d}{d\tau}\,q=iz\,\frac{d}{dz}\,q=1+
\frac {1} {p _ +} \,\sum _ {n\neq 0} \,
\alpha ^ {(+)} _ n \, z ^ {-n}\,. \eqno (A.5)
$$
We draw attention to the fact that all the quantities
(A.1)-(A.5) should be considered to be formal series in the
elements of free associative commutative involutive algebra
\ $ {\cal A} ^ {(+)} $ \ with generator \ $ \{x _ +/p _ +,
\,\alpha ^ {(+)} _ m/p _ + ,\,$ $ \bar {\alpha} ^ {(+)}
_m/p _ + \, \} $. This assumption holds until any averages
are calculated relative to physical states. The coefficients
at the monomials relative to the generator of the algebra \
$ {\cal A} ^ {(+)} $ \ in the expansions of (A.1)-(A.5)
are final polynomials in \ $z $ \ and \ $z ^ {-1} $.
Therefore, the integrals in (A.3) and (A.4) are defined
correctly. Thereby the matrix elements of the matrixes (A.3)
and (A.4) belong to the algebra \ $ {\cal A} ^ {(+)} $.

From (A.2) we have:
$$
\var_C \ln {u (z)} = \var_C \ln {z} -\var_C \,\left (\, \sum _ {n\ne 0} \,
 \frac{1}{n}\,\frac{\alpha^{(+)}_n}{p_+}\,z^{-n}\,\right)\,.
$$
Here \ $ \var_C F (z) $ \ means the change of the function \
$F (z) $ \ (generally ambiguous around the contour \ $C $)
for a single counterclockwise circuit around the contour $C$
. For this definition the second term on the right-hand side
of the last equality makes no contribution, and we have
$$
\var_C \ln {u (z)} = \var_C \ln {z} =2\pi \, i\,. \eqno (A.6)
$$
From (A.2) and (A.5) we find:
$$
 \frac{du(z)}{dz}=\frac{u(z)}{z}\,\sum^{+\infty}_{n=
- \infty}\,\frac{\alpha^{(+)}_n}{p_+}\,z^{-n}\ne 0\,. \eqno (A.7)
$$
The last inequality is a corollary of the fact that in
algebra \ $ {\cal A} ^ {(+)} $ \ there are no relations
except for what follow from its commutativeness.

Image of the contour \ $C $ \ in algebra \ $ {\cal A} ^
{(+)} $ \ under the mapping (A.2) is denoted by \ $C ^ * $.

The mapping (A.2) can be inverted. This is a consequence of
inequality (A.7). This inversion implies that there is an
analytic function \ $z (u) $ \ of the variable \ $u $ which
inverts the equation (A.2) into an identity. The function \
$z (u) $ \ is a formal series of the variables \ $ \{\,
\alpha ^ {(+)}/p _ + \, \} $. Equation (A.2) may be inverted
by an iterative method in powers of the variables \ $ \alpha
^ {(+)}/p _ + $. Let us assume that
$$
z (u) =z ^ {(0)} (u) +z ^ {(1)} (u) + \ldots \,
$$
where
$$
 z^{(i)}(u)=\sum^{+\infty}_{n=-\infty}\,z^{(i)}_n\,u^{-n}
$$
and \ $z ^ {(i)} _n $ \ are homogeneous functions of the
variables \ $ \alpha ^ {(+)}/p _ + $ \ to the power \ $i $.
Using equation (A.2) we obtain:
$$
z ^ {(0)} (u) =u \, e ^ {-i\delta} \,, \ \
\delta =\frac {x _ +} {2p _ +} \,,
$$
$$
z ^ {(1)} (u) =u \, e ^ {-i\delta} \,
\sum _ {n\ne
 0}\,\frac{\alpha^{(+)}_n}{np_+}\,(e^{-i\delta}u)^{-n}\,, \
 \
$$
$$
z ^ {(2)} (u) =u \, e ^ {-i\delta} \,
\left\{\frac {1} {2} \,
\left [\,\sum _ {n\ne
  0}\,
  \frac{\alpha^{(+)}_n}{np_+}\,(e^{-i\delta}u)^{-n}\,\right]^2
  - \right.
$$
$$
\left. -\left [\, \sum _ {n\ne
  0}\,\frac{\alpha^{(+)}_n}{p_+}\,(e^{-i\delta}u)^{-n}\,\right]\,
\left [\, \sum _ {m\ne
 0}\,\frac{\alpha^{(+)}_m}{mp_+}\,
 (e^{-i\delta}u\,)^{-m}\,\right]\,\right\}\,,
                                        \eqno (A.8)
$$
and so on. Equation (A.2) uniquely determines each
successive term \ $z ^ {(i)} (u) $ \ in terms of the
previous ones.

From (A.6) it follows also that for a single
counterclockwise circuit of the variable \ $u $ \ around the
circuit \ $C ^ * $ \, the variable \ $z $ \ goes once
counterclockwise around the contour $C$ . From this it
follows that a closed integral over the variable \ $z $ \
along the circuit \ $C $ \ may be converted into a closed
integral over the variable \ $u $ along the circuit \ $C ^ *
$ and conversely. In accordance with (A.7) and (A.5) we have
$$
\frac {du} {u} = \frac {dz} {z} \, \left\{\, 1 + \sum _ {n\ne
 0}\,\frac{\alpha^{(+)}_n}{p_+}\,z^{-n}\,\right\}=
\frac {dz} {z} \, \dot {q}\,. \eqno (A.9)
$$
The matrices which are inverse of (A.3) and (A.4) are
represented most simply in the form
$$
{ \cal M} ^ {-1} _ {n, \, l} = \frac {1} {2\pi
   i} \, \oint _ {C ^ *} \,\frac {du} {u} \, u ^ {-l} z^n\,,
   \eqno (A.10)
$$
$$
{}^F{\cal M}^{-1}_{n, \, l} = \frac{1}{2\pi
 i}\,\oint_{C^*}\,\frac{du}{u}\,u^{-l}z^n\,(\dot{q}\,)^{-1/2}\,.
 \eqno (A.11)
$$
In order to prove the equalities
\ $ {\cal M} \, {\cal M}^{-1} = {\cal M}^{-1} \, {\cal M} =1 $, \
 $ {}^F {\cal M} \, {}^F {\cal M}^{-1} = $ $ {}^F {\cal M}^
{-1} \, {}^F {\cal M}=1\,, $ we use the identities
$$
 \left(z\,\frac{d}{dz}\,\right)^s\,f(z,\,u(z)\,)=\frac{1}{2\pi
 i}\,\oint_C\,\frac{dz_1}{z_1}\,f(z_1,\,u(z_1)\,)\,
 \sum^{+\infty}_{n=-\infty}\,n^s\,\left(\frac{z}{z_1}\right)^n\,,
$$
$$
\left (u \,\frac {d} {du} \, \right) ^s \,
f (u, \, z (u) \,) = \frac {1} {2\pi i} \,
 \oint _ {C ^ *} \,\frac {du_1} {u_1} \, f (u_1, \, z (u_1)
\,) \,
 \sum^{+\infty}_{n=-\infty}\,n^s\,\left(\frac{u}{u_1}\right)^n\,,
 \
   s=0, \, 1, \ldots \eqno (A.12)
$$
In (A.12) the function \ $f (z, \, u) $ \ is expanded in a
Laurent series in terms of its arguments and the integration
goes in the counterclockwise direction. The identities
(A.12) are based on the obvious equalities
$$
 \oint_C\,\frac{dz}{z}\,z^n=\oint_{C^*}\,\frac{du}{u}\,u^n=2\pi
 i \,\delta_n\,.
$$
As a result of (A.6) and (A.12) for \ $s=0 $ \ we have:
$$
\sum ^ {+\infty} _ {l =
-\infty} \, {\cal M}_{m, \, l} \, {\cal M} ^ {-1} _ {l, \, n} =
\frac {1} {2\pi i}\,\oint_{C^*}\,\frac{du_1}{u_1}\,u^{-n}_1\cdot
\frac {1} {2\pi i} \, \oint_C \,\frac {dz} {z} \, u^m \,
 \sum^{+\infty}_{l=-\infty}\,\left(\frac{z_1}{z}\right)^l=
$$
$$
= \frac {1} {2\pi
i}\,\oint_{C^*}\,\frac{du_1}{u_1}\,u^{m-n}_1=\delta_{m-n}\,.
  \eqno (A.13)
$$
In (A.13) it is meant that $u=u (z) $ and $z_1=z (u_1) $.
Thus, the equality \ $ {\cal M} \, {\cal M} ^ {-1} =1 $ \ is established.
The relations \ $ {\cal M} ^ {-1} \, {\cal M} =1 $,
\ $ {} ^F {\cal M} \, {} ^F {\cal M} ^ {-1} =1 $,
and\ $ {} ^F {\cal M} ^ {-1} \, {} ^F {\cal M} =1 $. are
established similarly.

Formulas (A.3) and (A.10) directly yield the relation
$$
l \, {\cal M} ^ {-1} _ {n, \, l} = n \, {\cal M} _ {-l, -n}
\,, \eqno (A.14)
$$
which is valid for all \ $l $ \ and \ $n $. From (A.4),(A.9)
and (A.11) we also obtain:
$$
{} ^F {\cal M} ^ {-1} _ {n, \, l} = {} ^F {\cal M} _ {-l,
-n} \,. \eqno (A.14 ')
$$
The nonzero commutators of \ $ \alpha ^ {(-)} _ m $ \ and \
$ \bar {\alpha} ^ {(-)} _m $ \ with
 \ $u (z) $ \ and \ $ \bar {u} (z) $ \ are obtained using
 (A.2) and (2.12):
$$
[ \,\alpha^{(-)}_m,\,u(z)\,]=-\frac{2}{p_+}\,z^m\,u(z)\,, \
\ m\neq 0 \,, \ \ \ [
\,\alpha^{(-)}_0,\,u(z)\,]=-\frac{1}{p_+}\,z^m\,u(z)\, \ \
 \eqno (A.15)
$$
and similarly for barred quantities.

From the definitions (A.3) and (A.4) and relations (A.15) it
follows that
$$
[ \, \alpha ^ {(-)} _ m, \, {\cal M} _ {n, \, l} \,] =
-\frac {2n} {p _ +} \, {\cal M} _ {n, \, l-m} \,,
$$
$$
[ \, \alpha ^ {(-)} _ m, \, {} ^F {\cal M} _ {n, \, l} \,]
=
-\frac {1} {2\pi i\,p_+}\,\oint_C\,\frac{dz}{z}\,z^{-l+m}\,u^n\,
( 2n\,(\dot{q}\,)^{1/2}+m\,(\dot{q}\,)^{-1/2}\,)\,, \ \
m\neq 0 \,,
$$
$$
[ \, \alpha ^ {(-)} _ 0, \, {\cal M} _ {n, \, l} \,] =
-\frac {n} {p _ +} \, {\cal M} _ {n, \, l} \,, \ \ \
[ \, \alpha ^ {(-)} _ 0, \, {} ^F {\cal M} _ {n, \, l} \,]
=
-\frac {n} {p _ +} \ {,} ^ F {\cal M} _ {n, \, l} \,. \eqno (A.16)
$$
The commutation relations (A.16) conserve their form if
barred quantities are substituted everywhere in (A.16). The
relations (A.16) exhaust all the nonzero commutators between
the values \ $ \alpha^{(-)}, \ \bar {\alpha}^{(-)}, $ \
 $ {\cal M}, \ \bar {\cal M}, $ \ $ {} ^F {\cal M}$, and $^F\bar {\cal M}, $.

Now with the help of (A.14) and (A.16) we obtain:
$$
[ \, \alpha ^ {(-)} _ m, \, {\cal M} ^ {-1} _ {n, \, l} \,] =
\frac {2n} {p _ +} \, {\cal M} _ {-l, \, - (m+n)} \,, \ \ \ m\neq 0
\,,
$$
$$
[ \, \alpha ^ {(-)} _ 0, \, {\cal M} ^ {-1} _ {n, \, l} \,] =
\frac {n} {p _ +} \, {\cal M} _ {-l, \, -n} =
\frac {l} {p _ +} {{\cal M} ^ {-1} _ {n, \, l}}\,. \eqno (A.17)
$$

In Section 3 the following formulas are used:
$$
\sum _ {q} \, {\cal M} ^ {-1} _ {m, \, l+q} \, {\cal
M} ^ {-1} _ {n, \, p-q} = {\cal M} ^ {-1} _ {(m+n), \,
(l+p)} \,,
                                              \eqno (A.18a)
$$
$$
\sum _ {q} \, q \,\left ({\cal M}^{-1} _ {m, \, l+q} \, {\cal
M} ^ {-1} _ {n, \, p-q} - {\cal M}^{-1} _ {n, \, l+q} \, {
\cal M} ^ {-1} _ {m, \, p-q} \, \right)
= (m-n) \, {\cal M} _ {- (l+p), - (m+n)}\,, \eqno (A.18b)
$$
$$
 \sum _ {q} \, q^2 \,\left ({\cal M} ^ {-1} _ {m, \, l+q} \,
{\cal M} ^ {-1} _ {n, \, p-q} - {\cal M}^{-1} _ {n, \, l+q}
\, { \cal M} ^ {-1} _ {m, \, p-q} \, \right)
= (m-n) \, (p-l) {\cal M} _ {- (l+p), - (m+n)}\,. \eqno (A.18c)
$$
Using the formulas (A.10) and (A.12) for \ $s=1 $ \ we can
obtain the following relation:
$$
\sum ^ {+\infty} _ {q =-\infty} \, q \,
{\cal M} ^ {-1} _ {m, \, l+q} \, { \cal
 M}^{-1}_{n,\,p-q}=\frac{1}{2\pi\,i}\,\oint_{C^*}\, du\cdot
u^{-p}\,z^n\,\frac{d}{du}(\,u^{-l}\,z^m\,)\,.
                                              \eqno (A.19)
$$
Since we are interests only in the antisymmetric component
of the relation (A.19) in terms of the indices \ $m $ \ and
\ $n $ \, on the right-hand side it is possible to remove \
$u ^ {-l} $ \ from beneath the derivation sign. Taking into
account also that \ $ (dz/du) \, du=dz $ we obtain (A.18b).
The remaining formulas (A.18) are proved similarly.

Let's indicate also the following formulas:
$$
\sum ^ {+\infty} _ {m =-\infty} \,
{\cal M} ^ {-1}_{m, s} \, {\cal M} ^ {-1} _ {n-m, r} =
 \frac{1}{p_+}\,\sum^{+\infty}_{m=-\infty}\,
\alpha ^ {(+)} _ {n-m} \, {\cal M}^{-1}_{m, r+s}\,, \eqno(A.20)
$$
$$
\sum _ {l} \, l \,\left ({} ^F {\cal M} ^ {-1} _ {n-l, \, p} \, {}^F {\cal
M} ^ {-1} _ {l, \, q} - {} ^F {\cal M} ^ {-1} _ {n-l, \, q} \, {}^F {\cal
M} ^ {-1} _ {l, \, p} \, \right)
 =\frac{(q-p)}{p_+}\,\sum_l\,\alpha^{(+)}_l\,
 {\cal M}^{-1}_{(n-l), (p+q)}
\eqno (A.21)
$$
which are derived using the relations (A.9),(A.10), and (A.12).
\centerline {}
\centerline {}
\pagebreak

\centerline {References}
\centerline {}
\centerline {}

\begin {itemize}
\item [1].
R. Jackiw, E-print archive, gr-qc/9612052.
\end {itemize}
\begin {itemize}
\item [2].
E. Benedict, R. Jackiw, H.-J. Lee, Phys. Rev. {\bf {D54}} (1996) 6213.
\end {itemize}
\begin {itemize}
\item [3].
D. Cangemi, R. Jackiw, B. Zwiebach, Ann. Phys. (N.Y). {\bf {245}} (1996)
408;
D. Gangemi and R. Jackiw, Phys. Lett. {\bf {B337}},
271 (1994); Phys. Rev. {\bf {D50}}, 3913 (1994); D. Amati, S. Elitzur and E.
Rabinovici, Nucl. Phys. {\bf {B418}}, 45 (1994); D. Louis-Martinez, J
Gegenberg and G. Kunstatter, Phys. Lett. {\bf {B321}}, 193 (1994); E.
Benedict, Phys. Lett. {\bf {B340}}, 43 (1994); T. Strobl, Phys. Rev. {\bf
{D50}}, 7346 (1994).
\end {itemize}
\begin {itemize}
\item [4].
S.N.Vergeles, Zh.Eksp.Teor.Fiz. {\bf {113}}, (1998)1566.
\end {itemize}
\begin {itemize}
\item [5].
O. Andreev, Phys.Rev. {\bf {D57}} (1998) 3725.
\end {itemize}
\begin {itemize}
\item [6].
S.N. Vergeles, E-print archive hep-th/9906024.
\end {itemize}
\begin {itemize}
\item [7].
P.A.M. Dirac, Lectures on quantum mechanics.
Yeshiva University New York. 1964.
\end {itemize}
\begin {itemize}
\item [8].
P.A.M. Dirac. Lectures on quantum field theory. Yeshiva University,
 New York, 1967.
\end {itemize}
\begin {itemize}
\item [9].
E. Del Giudice, P.Di Vecchia, S. Fubini. Ann. Phys. {\bf {70}} (1972) 378.
\end {itemize}
\begin {itemize}
\item [10].
M.B.Green, J.H. Schwarz, E. Witten. Superstring Theory. Cambridge
University Press, 1987.
\end {itemize}
\begin {itemize}
\item [11].
 P.A.M. Dirac. The principles of quantum mechanics. Oxford 1958.
\end {itemize}

\end {document}